\documentclass[12pt, preprint, longabstract,usegraphicx]{aastex}
\usepackage{lscape}

\begin{document}
\title{Probing the Galactic Potential with Next-Generation Observations of Disk Stars}


\author{
T.~Sumi\altaffilmark{1},
K.V.~Johnston\altaffilmark{2},
S.~Tremaine\altaffilmark{3},
D.N.~Spergel\altaffilmark{4} \&
S. R.~Majewski\altaffilmark{5}
}

\altaffiltext{1}{Solar Terrestrial Environment Laboratory, Nagoya University, Nagoya, 464-8601, Japan\\ e-mail: {\tt sumi@stelab.nagoya-u.ac.jp}}
\altaffiltext{2}{Department of Astronomy, Columbia University, New York, NY 10027, USA\\ e-mail: {\tt kvj@astro.columbia.edu}}
\altaffiltext{3}{Institute for Advanced Study, Princeton, NJ 08540, USA,\\ e-mail: {\tt tremaine@ias.edu}}
\altaffiltext{4}{Princeton University Observatory, Princeton, NJ 08544-1001, USA,\\ e-mail: {\tt dns@astro.princeton.edu}}
\altaffiltext{5}{Department of Astronomy, University of Virginia,
Charlottesville, VA 22904-4325, USA \\ e-mail: {\tt srm4n@virginia.edu}}
\begin{abstract}
Our current knowledge of the rotation curve of the Milky Way is 
remarkably poor compared to other galaxies, limited by the combined 
effects of extinction and the lack of large samples of stars with good 
distance estimates and proper motions.  Near-future surveys promise a dramatic improvement in 
the number and precision of astrometric, photometric, and spectroscopic 
measurements of stars in the Milky Way's disk.  We examine the impact of such 
surveys on our understanding of the Galaxy by ``observing'' particle 
realizations of nonaxisymmetric disk distributions orbiting in an axisymmetric 
halo with appropriate errors and then attempting to recover 
the underlying potential using a Markov Chain Monte Carlo (MCMC) approach.  
We demonstrate that the azimuthally averaged gravitational force field in the Galactic plane---and
hence, to a lesser extent, the Galactic mass distribution---can be tightly constrained
over a large range of radii using a variety of types of surveys so long as the 
error distribution of the measurements of the parallax, proper motion, and radial 
velocity are well understood and the disk is surveyed globally.
One advantage of our method is that the target stars can be selected nonrandomly
in real or apparent-magnitude space to ensure just such a global sample without biasing the results. 
Assuming that we can always measure the line-of-sight velocity of a star with
at least 1 km\,s$^{-1}$ precision, we demonstrate that the force field can be
determined to better than $\sim$1\% for Galactocentric radii in the range $R=4-20$ kpc
using either: (1) small samples (a few hundred stars) with very accurate trigonometric 
parallaxes and good proper-motion measurements (
uncertainties $\delta_{p, {\rm tri}} \lesssim 10$ $\mu$as and 
$\delta_{\mu} \lesssim 100$ $\mu$as\,yr$^{-1}$ respectively);
(2) modest samples ($\sim 1000$ stars) with good indirect parallax estimates (e.g.,
uncertainty in photometric parallax $\delta_{p,{\rm phot}}\sim$ 10\%-20\%) and good proper-motion measurements
($\delta_{\mu} \sim 100$ $\mu$as\,yr$^{-1}$);
or (3) large samples ($\sim10^4$ stars) with good indirect parallax estimates and lower accuracy 
proper-motion measurements ($\delta_{\mu} \sim$ 1 mas\,yr$^{-1}$). We conclude that near-future surveys,
like SIM Lite, Gaia, and VERA,
will provide the first precise mapping of the gravitational force field in the region of the Galactic disk.
\end{abstract}

\keywords{
dark matter -- Galaxy:disk -- Galaxy:fundamental parameters -- Galaxy: kinematics and dynamics -- methods: data analysis -- surveys
}

\section{Introduction}
\label{sec:introduction}

Observations of the motions of stars and gas in galaxies tell us that they 
contain many times more mass in encompassing dark matter halos
than in their stellar components \citep[e.g.,][]{kent87}.
However, exactly how this dark matter is actually distributed 
in galaxies is still of some debate.
For example, while simulations of cold dark matter halos forming in an expanding 
universe seem to generally converge on a density distribution that can
be represented by a universal formula \citep{nfw96,nfw97},
the shape and radial profile of the inner parts of dark matter halos are still uncertain
\citep[see discussion in][]{navarro04,hayashi04}.

Of course, baryons are expected to complicate the elegant simplicity of the picture of dark matter halos painted by pure $N$-body simulations.
Gas radiates away energy to sink toward the centers of the dark matter halos where it can 
contribute significantly to the gravitational potential.
This process can cause the background dark matter halo
to contract further in response \citep[as reviewed in][]{gnedin04}
and evolve from triaxial to more spherical in shape \citep{dubinski94,katzantzidis04,bailin05,abadi09}.
On the other hand, stellar bars at the centers of galaxies can transfer angular
momenta to their host halos, flattening their  central density cusps \citep{sellwood06}.
The decay of satellite galaxies and substructure can also flatten the
 central density cusps.

Ultimately, we want to be able to distinguish between
dark and luminous contributions to the distribution of matter 
throughout galaxies. 
Stellar disks provide some of the cleanest probes of matter distributions, 
with stars moving on near circular orbits. 
Nevertheless, there remains the tricky problem of decomposing a disk galaxy 
potential into disk, bulge and halo contributions in order to isolate the 
form of the dark matter distribution.
One approach to this dilemma has been to look at low-surface-brightness galaxies, which
are expected to be dominated by dark matter,
yet even in these cases the results have been controversial and ambiguous \citep[see][for two opposing views]{deblok05,hayashi04}

It is striking that our own Milky Way galaxy
 has as yet contributed 
little to these debates.
After all, this is the one galaxy we can expect to 
study star-by-star with very high resolution in three-, four- 
or even six-dimensional phase space.
So far, three effects have hampered these ambitions:
first, our lack of accurate distance measurements to stars; 
second, our lack of accurate proper motions of stars;
and third, our inability to see across the Galactic disk because of dust absorption.
Because of these we do not yet have  the solar circular speed to better than 10\%,
the disk scale length to better than 20\%, or an accurate assessment of our own Galaxy's rotation 
curve beyond the solar circle \citep[see][]{olling98}. We have only fairly recently 
become convinced of the barred nature of the Milky Way \citep{bli91,weinberg92}
and are unsure whether we live in a flocculent or grand design spiral \citep{quillen02}.

The {\it Hipparcos} Space Astrometry Mission revolutionized our understanding of
the solar neighborhood
by compiling 1 milliarcsec level astrometry of 120,000 stars.
Using this data \citet{creze98} and \citet{holmberg00,holmberg04} measured the 
local matter density in the disk (the Oort limit) to be $\sim 0.1 M_{\sun}$pc$^{-3}$, 
a value that leaves little room for any significant contribution from disk dark matter.
Such an explicit decomposition of baryonic and dark matter contributions to 
a disk potential is impossible in external galaxies. 
\citet{flynn06} used these results to estimate the local surface mass-to-light ratios ($M/L$) 
for the Galactic disk of $(M/L)_V =1.5\pm 0.2\, M_\sun L_\sun^{-1}$ and inferred that the Milky Way is 
under-luminous by about 1$\sigma$ with respect to the Tully--Fisher 
relation; if the rotation speed announced by \citet{reid09} is correct this 
discrepancy is even more significant.  While these studies demonstrate the 
importance of large-scale, systematic Galactic studies to understanding galaxies in more detail,
Hipparcos' distance horizon was about 100 pc (distances of 10\% accuracy) so it
could not map the distribution of the mass in the Galaxy beyond the solar neighborhood. 

Three innovations in observations promise to dramatically
improve our understanding of the phase-space structure of our Galactic disk: 
(1) 
large-scale photometric surveys, both existing (the Two Micron All Sky Survey (2MASS) and the Sloan Digital Sky
Survey) and planned (PanSTARRS and LSST), together with methods of
deriving accurate photometric parallaxes for stars in these surveys
\citep{maj03,juric08}.
(2) high-precision (few to 10's of $\mu$as) astrometry from 
radio observations of masers e.g., VERA \citep[]{hon00}, 
VLBA \citep[]{reid08,hachisuka09} and the European VLBI Network \citep[]{rygl08}
and optical observations of stars 
(NASA's SIM Lite---Space Interferometry Mission Lite and  
ESA's GAIA---Global Astrometric Interferometer for Astrophysics, see
\citealt{unwin07}\footnotemark\footnotetext{\tt This actually presents about SIM PlanetQuest 
instead of SIM Lite}; \citealt{per02});
and 
(3) large-scale, high-resolution spectroscopic surveys, such as the ongoing 
Radial Velocity Experiment \citep[RAVE;][]{steinmetz06} and the SEGUE project of 
the Sloan Digital Sky Survey \citep{beers04} as well as  the planned
Apache Point Observatory Galactic Evolution Experiment \citep[APOGEE;][]{allende08}, HERMES instrument for the Anglo Australian Telescope and 
Wide Field Multi-Object Spectrograph (WFMOS) for the Gemini telescope.  It is clear that any or 
(better yet) all of these advances will significantly improve our knowledge of the Galaxy.  
What is unclear is the relative contribution of each type of survey: how uncertain and/or biased will 
our mass estimates be if one (or more) dimensions of phase space remain unmeasured? 
How far across the Galactic disk do we need to probe in order to construct
its rotation curve confidently?  To what extent can measurement errors be compensated for 
by using large numbers of stars?
Our study represents a first step toward addressing these questions.

Here we describe a general method to recover the underlying potential of the Galaxy
from photometric, astrometric, and spectroscopic surveys of disk stars (Section \ref{sec:recover}).
We test our method by constructing nonaxisymmetric particle
disks orbiting in a given potential, 
simulating observations of these particles with varying degrees of
accuracy, sample size, and disk coverage (Section \ref{sec:disks})
and examining how well the underlying potential can be measured.
We present the results of applying the recovery routine
to our ``observed'' data sets in
Section \ref{sec:results},
discuss the implications of these
results for future surveys in  Section \ref{sec:discussion},
and summarize our conclusions in Section \ref{sec:conclusion}.

\section{Methods}
\label{sec:method}

\subsection{Particle Disk Realizations}

This paper focuses on how astrometric
determinations of the motions of disk stars in the Galaxy can
best be utilized to measure the total potential in which they are
moving: we neither attempt to disentangle the disk and halo
contributions to the potential nor model motions perpendicular to the
Galactic plane, although the methods that we describe in this
 paper can easily be extended to these tasks.
Moreover, we do not address the use of radial velocity-only surveys.
We explore the power of astrometric measurements to measure the 
Galactic potential by using an approximate, parameterized kinematical model to generate a stellar sample,
adding observational errors and then attempting to recover the parameters of the model. 

\label{sec:disks}

The positions and motions of particles in our models
are generated from analytical formulae derived for 
axisymmetric disks 
perturbed by spiral arms.  We stress that our
use of approximate analytical formula (e.g., from epicycle theory) does not
compromise the validity of our results so long as the {\em same} approximate
formulae used to generate the particle realization are used to recover the
parameters of the model from observations of the realization.
In other words, the tests described in this paper provide an accurate
assessment of the validity of our method so long as accurate physics
is used when analyzing the real data. Of course, a potential problem 
for applying this method---or any method based on parameterized models---is
that the results may be misleading if the parameterized models are not
an accurate description of the real Galaxy.

The total potential is written as a sum of
axisymmetric $\Phi$ and spiral arm $\Phi_s$ terms
\begin{equation}
  \Phi_{\rm total}=\Phi(R) +\Phi_s(R,\phi,t)
\label{eq:potential}
\end{equation}
where $R$ is the Galactocentric radius and $\phi$ is the azimuthal angle in
the disk, measured from the Sun--Galactic center (GC) line and increasing in the 
same direction as Galactic rotation.

The particles are assumed to be drawn from an underlying axisymmetric
distribution of number density $\Sigma_{\rm sym}(R)$, whose response to the
spiral arm potential perturbation $\Sigma_s(R,\phi,t)$ is calculated
in the linear regime, to give a total number density:
\begin{equation}
  \Sigma=\Sigma_{\rm sym}(R) +\Sigma_s(R,\phi,t).
\label{eq:Sigma_total}
\end{equation}

The motions of the particles in the underlying potential are chosen to maintain the
number density distribution: the
mean radial ($\overline{v}_R$) and azimuthal ($\overline{v}_\phi$) speeds are given by
\begin{eqnarray}
      \overline{v}_R&=& \overline{v}_{R, \rm sym}(R) + v_{R s}(R,\phi,t),
\label{eq:vR} \\
      \overline{v}_\phi&=& \overline{v}_{\phi, \rm sym}(R) + v_{\phi s}(R,\phi,t),
\label{eq:vPhi}
\end{eqnarray}
where  ($\overline{v}_{R, \rm sym}$, $\overline{v}_{\phi, \rm sym}$) are the mean radial 
and azimuthal speeds, respectively, set by the gravitational potential of the axisymmetric disk,
and ($v_{Rs}$, $v_{\phi s}$) are additional perturbations to the mean due to spiral structure (see Equations \ref{eq:vr}-\ref{eq:B} below).

The formulae, adopted functional form and parameters for axisymmetric
and spiral arm terms in Equations (\ref{eq:potential}) -
(\ref{eq:vPhi}) are described in Sections \ref{sec:axis} and
\ref{sec:spiral}, respectively.  The dashed lines in Figure \ref{fig:spiral} 
show the results for our 
standard model (hereafter, the INPUT model), and the dots show the
velocities for a sample of particles in the range
$\phi=\pm\pi/16$ radians generated from this model without errors.

\subsubsection{Axisymmetric potential, number density and motions}
\label{sec:axis}

The realized disks are zero-thickness and exponential in Galactocentric radius $R$:
\begin{equation}
      \Sigma_{\rm sym}(R) = \Sigma_0  \exp\left( -\frac{R}{h}\right).
\label{eq:nexp}
\end{equation}
where $\Sigma_0$ is the central value and 
$h=3.0$ kpc is the scale length of the disk number density.

We work in terms of a spherical mass distribution for simplicity, even though the
actual mass distribution is certainly flattened, because we are only
modeling the potential in the disk midplane. 
The combined disk and halo mass distribution is represented by
a Hernquist function
(of total mass $M_h=10^{12}$ $M_{\odot}$ and scale length $a=22$ kpc, \citealt{her90}),
\begin{equation}
        M(R)=\frac{M_hR^2}{(R+a)^2},
\label{mhalo}
\end{equation}
where $M(R)$ is the mass enclosed within radius $R$. The corresponding potential is
\begin{equation}
        \Phi(R)=\frac{-GM_h}{(R+a)}.
\end{equation}

The circular velocity in this potential 
is calculated via
\begin{equation}
v_{\rm circ}(R)=\sqrt{R{d\Phi \over dR}}=\sqrt{GM\over R}={\sqrt{GM_hR} \over (R+a)}.
\end{equation}
The radial velocity dispersion is assumed to follow
\begin{equation}
      \sigma^2_{R}(R) = 
           \sigma^2_{R,\sun} 
           \exp\left( \frac{R_0-R}{h_\sigma}\right),
\label{eq:sigr}
\end{equation}
where $\sigma_{R,\sun}=25$ km\,s$^{-1}$ is the radial velocity dispersion at the Sun
(taken to be at $R_0=8$ kpc from the GC). Our potential recovery algorithm 
determines $h_\sigma$ independently of $h$, i.e., it does not assume that
$h_\sigma=h$. If the disk is self-gravitating, the shape of the velocity ellipsoid is 
independent of radius, and the disk thickness is independent of radius 
(as observed in external galaxies)
then we expect $h_\sigma = h$ (see details in \citealt{her93}).
Thus our input model assumes $h_\sigma=h=3.0$ kpc.
It is not certain that this assumption is true for the Galactic disk, since estimates for $h$
\citep[e.g.,][found $h=2-2.5$ kpc for the old disk stars]{ojha94}  
and $h_\sigma$ \citep[e.g.,][found $h_\sigma=4.37 \pm 0.32$ kpc in a study of old disk K-giants]{lewis89} 
are not same.

The azimuthal velocity dispersions $\sigma_\phi$ are assigned according to 
epicycle theory,
\begin{equation}
    \sigma_\phi(R)=\sigma_R\sqrt{\frac{\kappa^2}{4\Omega^2}},
\label{sigmaphi}
\end{equation}
where $\Omega=v_{\rm circ}/R$ is the angular velocity of a circular orbit at radius $R$ and $\kappa$ is the
 epicyclic frequency given by
\begin{equation}
      \kappa^2 (R)
=\frac{\partial^2 \Phi}{\partial R^2} 
               +\frac{3v_{\rm circ}^2}{R^2}
               =\frac{GM}{R^3}+\frac{G}{R^2}\frac{dM}{dR}.
\label{eq:kappa}
\end{equation}
The mean azimuthal motion $v_{\phi, \rm sym}$ due to the axisymmetric potential is given 
by the asymmetric drift equation,
\begin{equation}
\overline{v}_{\phi, \rm sym}^2(R)=v_{\rm circ}^2-\sigma_{\rm \phi}^2+\sigma_R^2+\frac{R}{\Sigma_{\rm sym}}
\frac{\partial (\Sigma_{\rm sym} \sigma_R^2)}{\partial R}
 =v_{\rm circ}^2-\sigma_{\rm \phi}^2+\sigma_R^2-R\sigma_R^2\left(\frac{1}{h_\sigma}+\frac{1}{h}\right),
\label{eq:drift}
\end{equation}
where the second equality was derived using Equations (\ref{eq:nexp}) and (\ref{eq:sigr}).
The mean radial motion $\bar{v}_{R, \rm sym}$ is zero.

\subsubsection{Spiral arms}
\label{sec:spiral}
The $m$-armed spiral potential is of the form
\begin{equation}
	\Phi_{S}(R,\phi,t)=\Phi_{a}(R) \cos \left[m(\phi-\Omega_{p}t)+c
              \log \left(\frac{R}{\rm 8\,kpc}\right) + \phi_0 \right]
\label{eq:phim}
\end{equation}
where $\Phi_{a}$ is the amplitude of the spiral potential, 
$\Omega_{p}$ is its pattern speed and 
$k=c/R$ is the radial wave-number which is related to the
pitch angle, $\theta$, by $\cot{\theta} = c/m$.
For the remainder of the discussion we consider the case of an $m=2$ spiral,
of amplitude $\Phi_{a2}(R)= 200({\rm km\,s^{-1}})^2$, 
$\Omega_{p}=1.4$ km\,s$^{-1}$ kpc$^{-1}$, pitch angle $\theta =
15^{\circ}$ and phase $\phi_0=105^{\circ}$. 
This potential perturbation would be generated by spiral arm of mass 
density $6.9 M_\sun$pc$^{-2}$ (from Equation 6.30 of \citealt{bt08}) or 
about 10\% of the observed local disk surface density.

Note that these parameters are specifically chosen so that any resonances
lie outside our survey region; in particular the pattern speed is much 
lower than typical (very uncertain) estimates of the  pattern speed in the
Galaxy (\citealt{deb02}).  Moreover, with this pattern
speed the entire observable disk lies inside the inner Lindblad resonance, a
region in which self-consistent spiral waves normally do not propagate. This
oversimplification will need to be addressed in future work.

The linear-theory prediction  
for the velocity response of the system to the imposed potential is given in the tight-winding or WKB approximation  by
(\citealt{bt08});
\begin{eqnarray}
      v_{Rs}(R,\phi,t)   &=& v_{Ra}(R) \cos [m(\phi-\Omega_{p}t)+c\log{R \over R_0} +\phi_0],
\label{eq:vr}\\
      v_{\phi s}(R,\phi,t)&=& v_{\phi a}(R) \sin [m(\phi-\Omega_{p}t)+c\log{R \over R_0} +\phi_0],
\label{eq:vphi}
\end{eqnarray}
where
\begin{eqnarray}
        v_{Ra}(R)&=&\frac{m(\Omega-\Omega_p)}{\Delta}k\Phi _a {\it F},\\
        v_{\phi a}(R) &=&-\frac{2B}{\Delta}k\Phi _a {\it F},
\label{eq:va}
\end{eqnarray}
\begin{equation}
        \Delta=\kappa^2 - [m(\Omega-\Omega_p)]^2,
\label{eq:delta}
\end{equation}
and  
\begin{equation}
        B= -\frac{1}{2}\left[ \frac{d(\Omega R)}{dR} \right] 
         = -\Omega-\frac{1}{2}R\frac{d\Omega}{dR}.
\label{eq:B}
\end{equation}
In the above equations $F$ is the reduction factor given by 
Equation (6-63) in \cite{bt08}
and the function $B(R)$ is equal to Oort's B constant at $R=R_0$.

The number density perturbation due to
the spiral potential of Equation (\ref{eq:phim}) is given by
\begin{equation}
        \Sigma_s(R,\phi,t) = \Sigma_a(R) \cos \left[m(\phi-\Omega_{p}t)+c\log{\left(R \over 8\,{\rm kpc} \right)} +\phi_0 \right],
\label{eq:na}
\end{equation}
where the amplitude of the response number density $\Sigma_a$ 
is given by Equation (6-59) in \cite{bt08}:
\begin{equation}
        m(\Omega-\Omega_p)\Sigma_a + k\Sigma_0 v_{Ra} =0.
\label{eq:Sigmaa}
\end{equation}

\subsubsection{Rendering and ``Observing'' the Particle Disks}
\label{sec:observing}

The stellar sample generated from our particle disks
simulates the collection of data by a targeted astrometric study, such
as might be performed with SIM Lite. In particular, the target stars are selected 
based on estimated photometric parallaxes, then the trigonometric parallaxes of 
these targets are observed with appropriate errors. 
We anticipate that our results will be
broadly applicable to global astrometric surveys (e.g., the GAIA
mission), as well as smaller samples currently available from ground-based surveys (e.g., VERA).
The error distribution function would differ in these cases, but we defer more
detailed discussion of these alternative applications to Section \ref{sec:discussion}.

We use the following steps to generate our simulated data.
\begin{enumerate}
\item
The stars in the disk are generated over a radius range from $R=0$ kpc to $25$ kpc,
distributed according to the number density distribution of Equation (\ref{eq:Sigma_total}).
These stars are viewed from a point moving (in
Galactic rest-frame coordinates) with radial and azimuthal speeds
$(v_{R,{\rm view}},v_{\phi, {\rm view}})=(v_{R,{\rm LSR}},v_{\phi, {\rm LSR}})
=(\overline{v}_R, \overline{v}_\phi)$
at $R_{\rm 0}=$ 8 kpc from the GC (i.e., at the Sun) and each star
is assigned a Galactic longitude $l$ and parallax $p$.
Here we assume ${\bf v}_{\rm view}={\bf v}_{\rm LSR} $
where ${\bf v}_{\rm LSR}$ is the velocity of the Local Standard of Rest, because the
peculiar velocity of the Sun relative to the LSR is known \citep{dehnen98}.
The proper motion $\mu$ and line-of-sight velocity $v_{\rm los}$ of the stars
are assigned based on
$(v_R, v_\phi)$  
drawn from Gaussians with means $(\overline{v}_R,\overline{v}_\phi)$
given by Equations (\ref{eq:vR}) and (\ref{eq:vPhi}) and dispersions
$(\sigma_R,\sigma_\phi)$ given by Equations (\ref{eq:sigr}) and (\ref{sigmaphi}) at each position.
\item
The stars' parallaxes are scattered about their true values 
to give observed photometric parallaxes $p_{\rm phot}$ by drawing from
a Gaussian distribution of mean $p$ and dispersion $\delta_{p, {\rm
  phot}}$.
This observational uncertainty is assumed to be 15\% (i.e.,
$\delta_{p, {\rm phot}} = 0.15p$) in most of the following work except in 
Section \ref{sec:sigdeltapphot} (see also details in Section \ref{sec:standard}).
This step is intended to mimic the effect of
drawing a sample from a set of standard candles, with absolute
magnitudes in a well-characterized range.
Errors in $l$ are negligible.
\item 
  In our recovery algorithm, we parameterize the potential  
  using characteristic 
masses $M_i\equiv M(R_i)$ defined at eight discrete radii $R_i$ uniformly 
  spaced from $R_1=4$ kpc to $R_8=20$ kpc (see Section \ref{sec:param} for details).
  In order to have adequate constraints on each $M_i$, sample stars are selected based on observed $p_{\rm phot}$ (i.e.,
  including the uncertainty introduced by the distribution of
  absolute magnitudes of the standard candles) so that there are equal numbers
  of stars in the seven bins between the $R_i$.
  We impose the following additional restrictions on the sample,
  while retaining the constraint that there should be equal
  numbers of stars in each bin. 
  (1)  Sample stars are selected only between
  Galactocentric radii of $R_1$ and 
  $R_{\rm max}=\hbox{min}(18\hbox{\,kpc}, R_8 - 0.17d)$, where
  $d$ is the distance from the Sun.
The $d$ dependence in $R_{\rm max}$ is introduced to minimize the number of stars that scatter into our sample from outside our intended survey region since this could result in a systematic bias in our estimate for $M_8$. The dependence is tuned to the scale of the uncertainties: $0.17d$ corresponds to the 1$\sigma$
  uncertainty in distance due to a 15\% error in photometric
  parallax.
(2) No sample stars have $d>20$ kpc, to avoid stars with large distance
  errors. 
(3) Stars
  behind the Galactocentric circle at $R_1$ (i.e., $l<l_{\rm min}=30^\circ$ and
  $d>8\hbox{\,kpc}\cos l_{\rm min}$) are also excluded because of
  the typically strong extinction. 
  The observed spatial distribution (i.e., using $p_{\rm phot}$ to find location in the disk)
  of a sample of stars selected in this manner from the exponential disk
  is shown as black dots in Figure \ref{fig:plotxy}.
  The true spatial distribution of this sample of stars is also 
  shown as gray filled circles.
    Note that because these target stars are
  selected based on $p_{\rm phot}$ rather than $p$, some fraction of
  our sample actually lies outside our intended survey region.
\item
The observed trigonometric parallax, proper motion and line-of-sight velocity
$(p_{\rm tri}$, $\mu_{\rm o}$, $v_{\rm los,o})$ for our sample are assigned by
drawing from Gaussian distributions of mean $(p,\mu,v_{\rm los})$, given in
step 1 above, and dispersion $(\delta_{p, {\rm tri}}, \delta_\mu,\delta_{\rm los})$
where $\delta_{p, {\rm tri}}$ and $\delta_{\rm los}$ are constant and $\delta_\mu$ depends on 
$\delta_{p, {\rm tri}}$ (see details in Section \ref{sec:standard}).
This step is intended to mimic a targeted astrometric mission with ground-based spectroscopic follow-up, 
observing the sample stars with integration times tailored to achieve a constant accuracy. 
The faintest (i.e., most distant and smallest $p$) stars in any
sample may have more accurate photometric parallaxes than trigonometric parallaxes.
\end{enumerate}

\subsection{Recovering the Underlying Potential}
\label{sec:recover}

The recovery program uses a Markov Chain Monte Carlo approach (MCMC,
as described in \citealt{gilks96,verde03} and summarized below in Section
\ref{sec:mcmc}) to find the maximum and the shape of
the likelihood function
\begin{equation}
L({\bf x})=\prod_{i=1}^{N} P(p_{\rm tri}^i,\mu_{\rm o}^i,v_{\rm los,o}^i|p_{\rm phot}^i,l^i,{\bf x}),
\label{eq:L}
\end{equation}
for a sample of size $N$ as the parameters of the model ${\bf x}$ are
varied.  Here $P(p_{\rm tri}^i,\mu_{\rm o}^i,v_{\rm los,o}^i|$ $p_{\rm phot}^i$,$l^i$,${\bf x})$
is the conditional probability (derived in Section \ref{sec:like}) of observing a
star to have trigonometric parallax, proper motion and line-of-sight velocity $(p_{\rm tri},\mu_{\rm
  o},v_{\rm los,o})$, given its observed photometric parallax $p_{\rm phot}$ and Galactic
longitude $l$, and underlying Galaxy model parameters ${\bf x}$ (see Section \ref{sec:param}).

$P$ represents the conditional probability of observing a star's
kinematical properties at a particular position in the disk.
Hence, although we
need to have an appropriate model of the intrinsic
stellar spatial distribution from which we are selecting the sample (in order to
understand the likelihood of finding a star of given $p_{\rm phot}$ and $p_{\rm tri}$ in
the initial random survey), we have complete freedom in specifying how
we select the stars in our targeted sample.  This means we can choose
to distribute our tracers to regions of the disk that we are most
interested in resolving. In our case (as noted in Section 
\ref{sec:observing}), 
we select equal numbers of
stars with observed Galactocentric radii (based on the photometric
parallax) in each of the seven bins between the $R_i$,
rather than simply taking a random sample, and this
allows us to explore the outer disk in greater detail.

\subsubsection{The Markov Chain method}
\label{sec:mcmc}

In a single step of a MCMC run, the likelihood $L_{i, {\rm prop}}$
(Equation \ref{eq:L}) is evaluated for the model parameters ${\bf
  x}_{i, {\rm prop}}$ proposed at step $i$ in a chain and compared
with $L_{i-1}$ from the previous step.  If $L_{i, {\rm prop}} > q L_{i-1}$, 
for a random number $q$ between 0
and 1, the proposed parameters will be adopted for this step (${\bf
  x}_i={\bf x}_{i, {\rm prop}}$). Otherwise the parameters
from the previous step (${\bf x}_i={\bf x}_{i-1}$) are kept. Proposed
parameters for the next step (${\bf x}_{i+1}$) are generated by adding
a vector of small changes to ${\bf x}_i$. These steps are accumulated
until they satisfy the convergence criteria outlined in
\cite{verde03}.

The beauty of the MCMC method is that the distribution of the accepted
steps follows the shape of the likelihood function in parameter
space. This property of the method means that the chains of steps
themselves can be exploited in two ways: first they can be used to
derive the optimal directions and sizes of steps for exploring
parameter space (i.e., to get better acceptance ratios and faster
convergence); and second, they can be used to determine best
values and confidence intervals for each parameter.

In this work, a ``test'' MCMC is first run with step sizes for parameter
changes estimated from simple intuition. 
The covariance matrix of these preliminary chains is then constructed, and  the
eigenvalues and eigenvectors of the matrix are used to estimate optimal size
and direction of parameter vectors of the steps made in the following
actual MCMC runs.  This refinement is particularly important when the
parameters have strong correlations (see Section \ref{sec:results}).

The ``best'' values of parameters presented in all figures and tables
are taken to be the mean ${\bf x}$ from the chains, weighted by the
likelihood. The 1-$\sigma$ error bars represent 68\% confidence
intervals.

\subsubsection{Estimating the likelihood of a given parameter set}
\label{sec:like}

The likelihood in Equation (\ref{eq:L}) is the product of factors
$P(p_{\rm tri},\mu_{\rm o},v_{\rm los,o}|p_{\rm phot},l,{\bf x})$ for each star in the sample, which is the
conditional probability of making observations of trigonometric parallax $p_{\rm tri}$, proper motion
$\mu=\mu_{\rm o}$ and line-of-sight velocity
$v_{\rm los}=v_{\rm los,o}$ given an observed photometric parallax $p_{\rm phot}$, along
Galactic longitude $l$,
\begin{eqnarray}
        P(p_{\rm tri},\mu_{\rm o},v_{\rm los, o}|p_{\rm phot},l,{\bf x}) 
        ={P(p_{\rm phot},p_{\rm tri},\mu_{\rm o},v_{\rm los,o}|l,{\bf x}) \over P(p_{\rm phot}|l,{\bf x})} \cr
        =\frac{P(p_{\rm phot},p_{\rm tri},\mu_{\rm o},v_{\rm los, o}|l,{\bf x})}
{\int_0^\infty dp_{\rm tri} \int_{-\infty}^\infty d\mu_{\rm o} \int_{-\infty}^\infty dv_{\rm los,o} P(p_{\rm phot},p_{\rm tri}, \mu_{\rm o},v_{\rm los,o}|l,{\bf x})
      }.
\label{eq:P}
\end{eqnarray}
The full probability distribution can be derived from the phase-space distribution function 
$f_{\bf x}$ which is the number of stars per unit velocity and per unit parallax, given by
\begin{equation}
\label{eq:fx}
        f_{\bf x}(p, v_{\rm tan},v_{\rm los},l)= f'_{\bf x}(p, v_{\rm tan},v_{\rm los},l)V(p)
\end{equation}
where $V(p)$ is the volume per unit parallax at
$p$ ($\propto p^{-4}$ in three-dimensional space and $\propto p^{-3}$ for our zero-thickness disk), and
$f'_{\bf x}$ is the number of stars per unit volume of phase space
predicted by the model with parameters ${\bf x}$:
\begin{equation}
\label{eq:fxspase}
        f'_{\bf x}(p, v_{\rm tan},v_{\rm los},l)=
                \Sigma(p,l) U_{\bf x}(v_{R},v_{\phi}).
\end{equation}
In the above equation $\Sigma(p,l)$ is the number density of stars per
unit area at ($p$, $l$) given by Equation (\ref{eq:Sigma_total}) and 
$U_{\bf x}$ is the number of stars per unit velocity predicted 
from the model parameters ${\bf x}$ at position $(p,l)$ given by 
\begin{equation}
\label{eq:Ux}
        U_{\bf x}(v_{R},v_{\phi})=g(v_{R}, \overline{v}_{R,\bf x}, \sigma_{R,\bf x}) 
                                  g(v_{\phi}, \overline{v}_{\phi,\bf x}, \sigma_{\phi,\bf x}).
\end{equation}
Here $g(y,\overline y,\sigma)$ denotes the value at $y$ of a Gaussian
distribution with mean $\overline y$ and dispersion $\sigma^2$ and the
quantities $(\overline v_{R,{\bf x}}, \overline v_{\phi,{\bf
x}})$ and $(\sigma^2_{R,{\bf x}}, \sigma^2_{\phi,{\bf x}})$ are the
mean velocity and velocity dispersion at parallax $p$ and longitude
$l$ from the model with parameters $\bf x$, given by Equations (\ref{eq:vR}), (\ref{eq:vPhi}),
(\ref{eq:sigr}), and (\ref{sigmaphi}) respectively.
Finally,  ($v_{\rm tan}=\mu/p,v_{\rm los}$) can be transformed to ($v_{R},v_{\phi}$) for given $(p,l)$, $v_{\rm view}$ and $R_{\rm 0}$.

In our experiment, we first observe $p_{\rm phot}$ for a random disk sample, with an error distribution $\epsilon(p_{\rm phot}|p)=g(p_{\rm phot},p,\delta_{p, \rm phot})$ about $p$. The distribution in $(p,p_{\rm phot})$ 
of stars is given by
\begin{equation}
        \Sigma(p,l)\epsilon(p_{\rm phot}|p)V(p).
\end{equation}
A subset of these 
stars is selected for our sample, with specified distribution $N_{\rm sample}(p_{\rm phot}, l)$ along
a given line of sight, which can be related to the selection function, $S(p_{\rm phot},l)$ (i.e.,
the probability of including a star in the survey at $(p_{\rm phot},l)$) by
\begin{equation}
        N_{\rm sample}(p_{\rm phot},l)= S(p_{\rm phot},l) \int_0^\infty \Sigma(p,l)\epsilon(p_{\rm phot}|p)V(p) dp.
\end{equation}
The total number of stars in the sample $N$ is given by summing the number of stars toward $l$,
\begin{equation}
        N_l=\int_0^\infty N_{\rm sample}(p_{\rm phot},l) dp_{\rm phot},
\end{equation}
and along all adopted lines of sight.

Hence the full distribution of properties of the sample will depend on its intrinsic
distribution in phase-space $f_{\bf x}$, filtered by $S(p_{\rm phot},l)$  and convolved 
with appropriate error distributions for the remaining observables,  $\epsilon(p_{\rm tri}|p)=g(p_{\rm tri},p,\delta_{p, \rm tri})$, $\epsilon(\mu_{\rm o}|\mu)=g(\mu_{\rm o},\mu,\delta_\mu)$ and  $\epsilon(v_{\rm los,o}|v_{\rm los})=g(v_{\rm los,o},v_{\rm los},\delta_{v_{\rm los}})$
(as outlined in Section \ref{sec:observing}).
The probability of finding a star in the survey is given by
\begin{eqnarray}
        &&P(p_{\rm phot},p_{\rm tri},\mu_{\rm o},v_{\rm los,o}|l,{\bf x}) \cr
                &&={S(p_{\rm phot},l) \over N_l} \int_0^\infty dp \int_{-\infty}^\infty {d\mu \over p} \int_{-\infty}^\infty dv_{\rm los} f_{\bf x}(p, \mu/p,v_{\rm los},l)\epsilon(p_{\rm tri}|p)\epsilon(p_{\rm phot}|p)\epsilon(\mu_{\rm o}|\mu) \epsilon(v_{\rm los,o}|v_{\rm los})\cr
                &&= {S(p_{\rm phot},l) \over N_l}  \int_0^\infty \Sigma(p,l)V(p)\epsilon(p_{\rm tri}|p)\epsilon(p_{\rm phot}|p)P_{p,\mu, v_{\rm los}}(p,\mu_{\rm o},v_{\rm los, o})dp,
\label{eq:intP}
\end{eqnarray}
where
\begin{eqnarray}
P_{p,\mu, {\rm v_{ los}}}(p,\mu_{\rm o},v_{\rm los, o}) &=&\int_{-\infty}^\infty \int_{-\infty}^\infty  U_{\bf x}(\mu/p, v_{\rm los}) \epsilon(\mu_{\rm o}|\mu)  \epsilon(v_{\rm los, o}|v_{\rm los})  \frac{d\mu}{p} dv_{\rm los}.
\label{eq:Pmu}
\end{eqnarray}

Substituting in Equation (\ref{eq:P}) gives
\begin{eqnarray}
        &&P(p_{\rm tri},\mu_{\rm o},v_{\rm los, o}|p_{\rm phot},l,{\bf x}) 
        ={P(p_{\rm phot},p_{\rm tri},\mu_{\rm o},v_{\rm los,o}|l,{\bf x}) \over P(p_{\rm phot}|l,{\bf x})}  \cr
        &&=\frac{\int_0^\infty \Sigma(p,l)V(p)\epsilon(p_{\rm tri}|p)\epsilon(p_{\rm phot}|p)P_{p,\mu, v_{\rm los}}(p,\mu_{\rm o},v_{\rm los, o})dp}
{\int_0^\infty dp_{\rm tri} \int_{-\infty}^\infty d\mu_{\rm o}  \int_{-\infty}^\infty  dv_{\rm los,o}  \int_0^\infty
\Sigma(p,l)V(p)\epsilon(p_{\rm tri}|p)\epsilon(p_{\rm phot}|p)P_{p,\mu, v_{\rm los}}(p,\mu_{\rm o},v_{\rm los, o})
       dp }.
\label{eq:P1}
\end{eqnarray}
Note that this expression is independent of our sample selection function $S(p_{\rm phot}, l)/N_l$: our analysis method leaves us free to choose a sample with arbitrary properties without biasing the results. 
Also, in the limit of negligible errors in the photometric parallax
(i.e., $\epsilon(p_{\rm phot}|p)=\delta(p_{\rm phot}-p))$,
 Equation (\ref{eq:P1}) simplifies to:
\begin{equation}
 P(p_{\rm tri},\mu_{\rm o},v_{\rm los, o}|p_{\rm phot},l,{\bf x}) 
        =	\epsilon(p_{\rm tri}|p_{\rm phot})P_{p,\mu, v_{\rm los}}(p_{\rm phot},\mu_{\rm o},v_{\rm los, o})
\label{eq:P1a}
\end{equation}
and our approach becomes insensitive to the underlying disk surface density distribution.

\subsubsection{Parameterizing the OUTPUT model}
\label{sec:param}

The model distribution function $f_{\bf x}$ (see Equation \ref{eq:fx}) is fully specified
by the spatial number density, mean velocities and velocity dispersions of stars as a 
function of position in the disk.

Our OUTPUT model can represent observations of axisymmetric motions using a
total of 12 free parameters (11 free parameters when we fix $R_0$) to 
describe both $f_{\bf x}$ and the
transformation from physical to observed coordinates: (1) $h$---the
scale length of the Galactic disk given in Equation  (\ref{eq:nexp}); 
(2) $\sigma_{R,\sun}$ and $h_\sigma$ (assuming the
functional form for the radial velocity dispersion given in Eq. \ref{eq:sigr}); 
(3) the masses $M_i$ ($i=1,2,...,8$) within Galactic radius, 
$R_i$ (from which the mass and its derivative $\partial M /\partial R$ at
any radius are found using cubic spline interpolation); and (4)
$R_0$.  We also analyze the sample assuming a known value $R_0=8$ kpc
in the following sections because we anticipate that it will
be well-constrained by other observations \citep[e.g.,
adaptive optics observations of stars around the black hole at the 
Galactic center;][]{eis03}.
The effect of allowing $R_0$ to be a free parameter or fixed is shown in 
Section \ref{sec:sigN} and \ref{sec:sigmaphi}.

All other
quantities needed are derived from these parameters using the
expressions given in Section \ref{sec:axis}.

Describing the spiral arms assuming $m=2$ requires an additional four parameters: the
constant $c$ which is related to the radial wave number, $k=c/R$; the pattern speed $\Omega_p$;
the arm's phase $\phi_0$ and the amplitude of the
spiral potential, $\Phi_{a}$.  The perturbations to the mean
velocities and number density are calculated using Equations (\ref{eq:vr}),
(\ref{eq:vphi}), and (\ref{eq:na}).

The 16 free parameters are
listed in Table \ref{tbl:param}, along with the INPUT values from which 
we derived our observed sample.

\section{Results}
\label{sec:results}

In this section, we explore the accuracy of results recovered by 
applying the MCMC method to simulated observations of our disk model, with
INPUT parameters given in Table \ref{tbl:param}.  
For our standard sample, we look at disk M-giant stars,
observed with fixed astrometric accuracies, as an example
of a plausible near-future experiment that might be performed by a mission 
such as SIM Lite (see Section \ref{sec:standard}).
M-giants are evolved, metal-rich stars and therefore typically relatively
young (several Gyrs in age); this makes them good dynamical tracers of the
mean disk potential.
We then go on to examine how our results depend 
on sample size (Section \ref{sec:sigN}), 
trigonometric and photometric accuracy 
(Sections \ref{sec:sigdeltatri} and \ref{sec:sigdeltapphot}) and 
disk coverage (Section \ref{sec:sigmaphi}).

\subsection{The standard sample}
\label{sec:standard}

Our photometric sample is assumed to be
composed of disk M-giants selected using the infra-red color in the
2MASS catalog.
We assume that the intrinsic scatter in 
the absolute magnitudes of the M-giants around a mean of 
$M_V=-2$ would result in 
a photometric parallax error of 15\%, i.e., $\delta_{p, {\rm phot}}=0.15 p$
\citep[as estimated by][]{maj03}.
Note that this scatter is in part due to
metallicity differences (\citealt{cho07}), and measuring  the metallicity
to about 0.3 dex would allow a 
parallax accuracy as good as $\sim$10\%.

We select $\sim$850 stars from our simulated M-giant survey to follow the
distribution outlined in Section \ref{sec:observing}, and
``observe'' them with a trigonometric parallax 
accuracy of $\delta_{p, {\rm tri}}=10$ $\mu$as.
With these parameters, the point at which photometric rather than trigonometric parallaxes
became more accurate would be at $p\sim 67$ $\mu$as ($\sim15$ kpc).
The proper motion accuracy is expected to scale as
\begin{equation}
\delta_\mu=[0.235 + 0.634\times \delta_{p, {\rm tri} }({\rm \mu as})]\,\mu {\rm as\,yr^{-1} }
\label{eq:mue}
\end{equation}
from the  
SIM Global Astrometry Time Estimator\footnotemark\footnotetext{\tt http://mscws4.ipac.caltech.edu/simtools/portal/login/normal/1?}. 
We assume a constant error in the line-of-sight velocity $\delta_{v, {\rm
    los}}$=1 km\,s$^{-1}$ can be achieved from ground-based
spectroscopic observations.

\begin{figure}
\epsscale{0.5}
\includegraphics[angle=-90,scale=0.6,keepaspectratio]{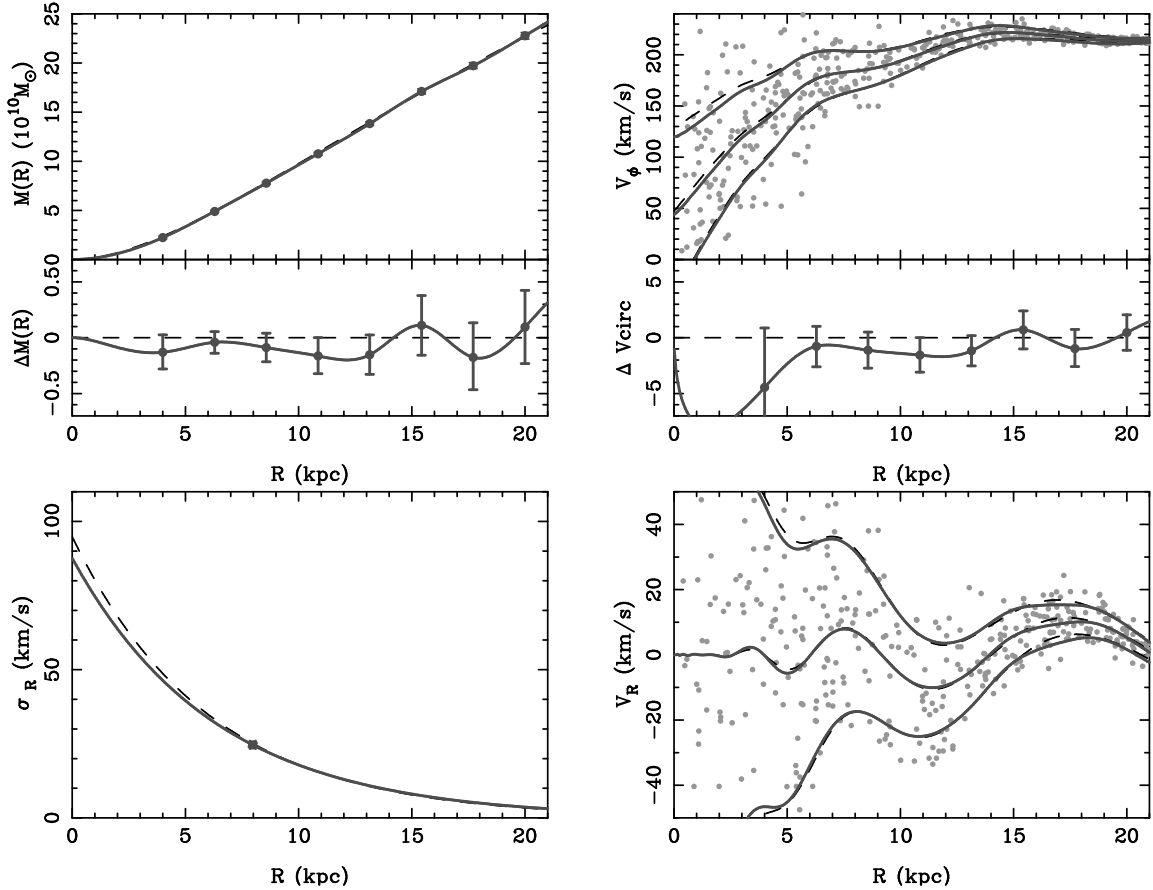}
\caption{INPUT (dashed lines) and recovered OUTPUT (solid lines with $\pm 1\sigma$ dispersion) models
as a function of the Galactocentric radius $R$.
The OUTPUT is derived for samples of $N=850$ stars, observed with assumed photometric parallax
and trigonometric parallax accuracies of $\delta_{p,{\rm phot}}=0.15p$ and $\delta_{p,{\rm tri}}=10$ $\mu$as. 
The model curves are interpolated from the mass parameters, $M_i$ 
(indicated for the OUTPUT as points with 1$\sigma$ error bars).
The oscillations in $v_R$ are due to spiral arms.
Top-left: mass within $R$, $M(R)$.
Middle left: residual from the INPUT mass, $\Delta M(R)$.
Top right: azimuthal velocity $v_{\phi}$. 
Middle right: residual from the INPUT circular velocity, $\Delta v_{\rm circ}$.
Bottom left: radial velocity dispersion $\sigma_{R}$, which is estimated from Equation
(\ref{eq:sigr}) using OUTPUT $\sigma_{R,\sun}$ and $h_\sigma$.
Bottom right: radial velocity $v_R$.
The values of $v_\phi$ and $v_R$ for stars with $\phi=\pm\pi/16$ are plotted as light gray dots.
In some plots the error bars are too small to be
resolved and the dashed and solid lines lie on top of one another.
We can recover the mass to within $\sim$2\% for $R$=4-20 kpc.
\label{fig:spiral}}
\end{figure}

Figure \ref{fig:spiral} illustrates the results of applying the MCMC recovery routine to this standard sample,
with the analytical estimates constructed
from INPUT and recovered OUTPUT parameters (listed in Table \ref{tbl:param})
shown as dashed and solid lines, respectively. 
The best fit values at interpolated points are plotted 
as filled circles, with 1-$\sigma$ error bars estimated directly from
the distribution of parameters in the MCMC.
The figure indicates that, with this level of accuracy, 
we can recover the mass distribution between 4 and 20 kpc to within
$\sim 2$\% using a sample size $\lesssim 10^3$.

\begin{figure}
\epsscale{1.0}
\includegraphics[angle=0,scale=0.7,keepaspectratio]{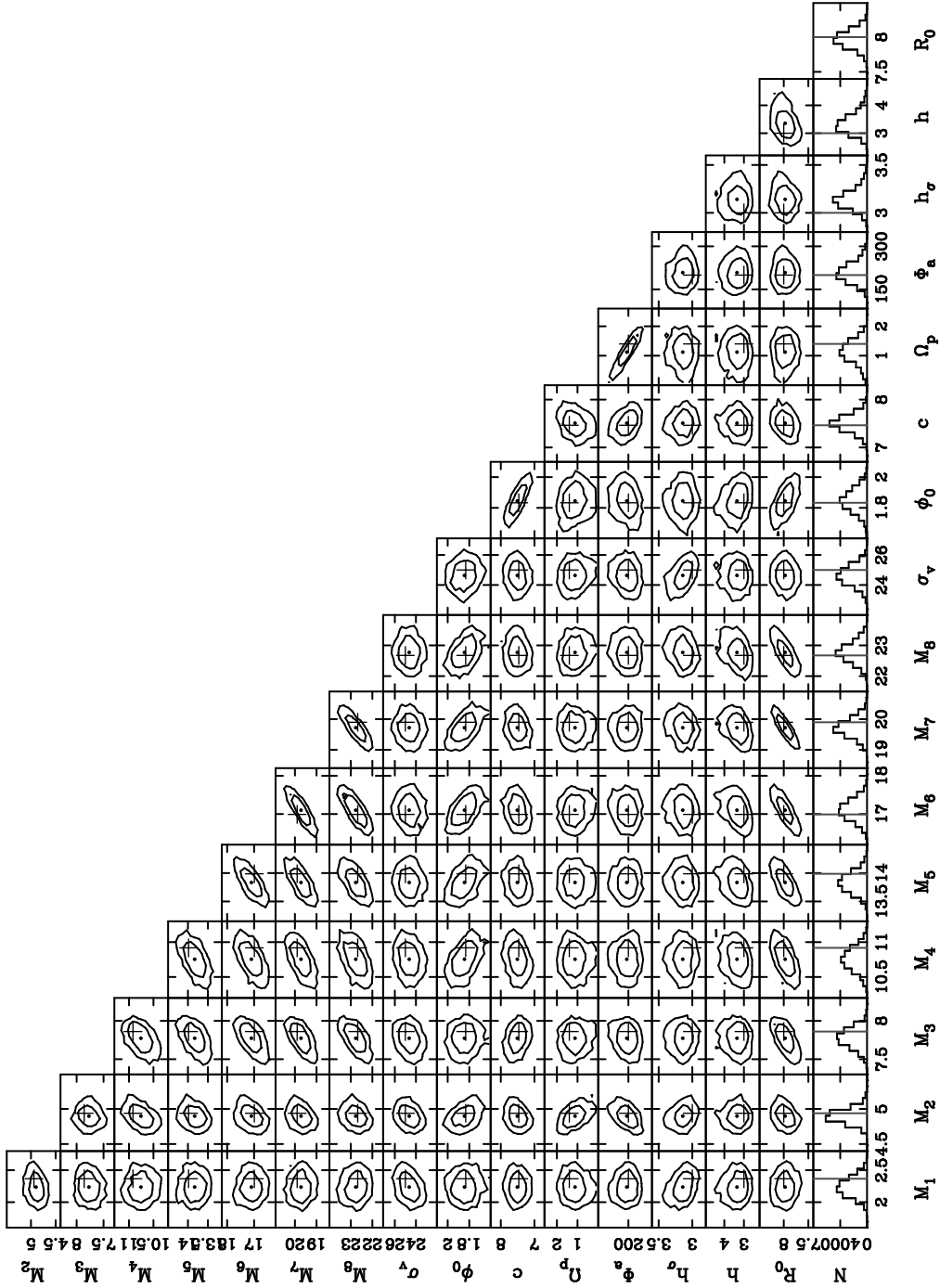}
\caption{ 
Likelihood distribution of parameters for the MCMC run shown in Figure \ref{fig:spiral},
with $N=850$, $\delta_{p,{\rm phot}}=0.15p$ and $\delta_{p,{\rm tri}}=10$ $\mu$as. 
The dots indicate the maximum likelihood values. 
The crosses indicate the INPUT values. 
The contours indicate 68\% and 95\% confidence intervals. The histograms represent the projected likelihood 
distribution for each parameter, where the vertical lines indicate INPUT values.
\label{fig:contour}}
\end{figure}

Figure \ref{fig:contour} shows the full likelihood distribution of parameters. 
There are strong correlations between 
(a) $M_i$'s and $R_0$: due to the relation $M =v_{\rm circ}^2R\,G^{-1}$; 
(b) $M_i$'s at large $R$: all $M_i$ correlate with $R_0$ simultaneously;
(c) $c$ and $\phi_0$: they define the phase of the spiral arms by Equations (\ref{eq:vr}) and (\ref{eq:vphi}); 
(d) $\Phi_a$ and $\Omega_p$: the amplitude of the mean velocity in the spiral arms
is related to them by Equations (\ref{eq:vr}) and (\ref{eq:vphi}).
As noted in Section \ref{sec:mcmc}, the steps in the MCMC were optimized using a preliminary run to take account 
of these correlations prior to our production runs.
However, these contours in correlated parameters become 
longer and banana-shaped
for small $N$, large observational errors and narrow coverage of samples in space 
(e.g., small $\phi_{\rm max}$ defined in Section \ref{sec:sigmaphi}), and under these
conditions the MCMC can fail to converge.

\begin{figure}
\epsscale{1.0}
\includegraphics[angle=-90,scale=0.7,keepaspectratio]{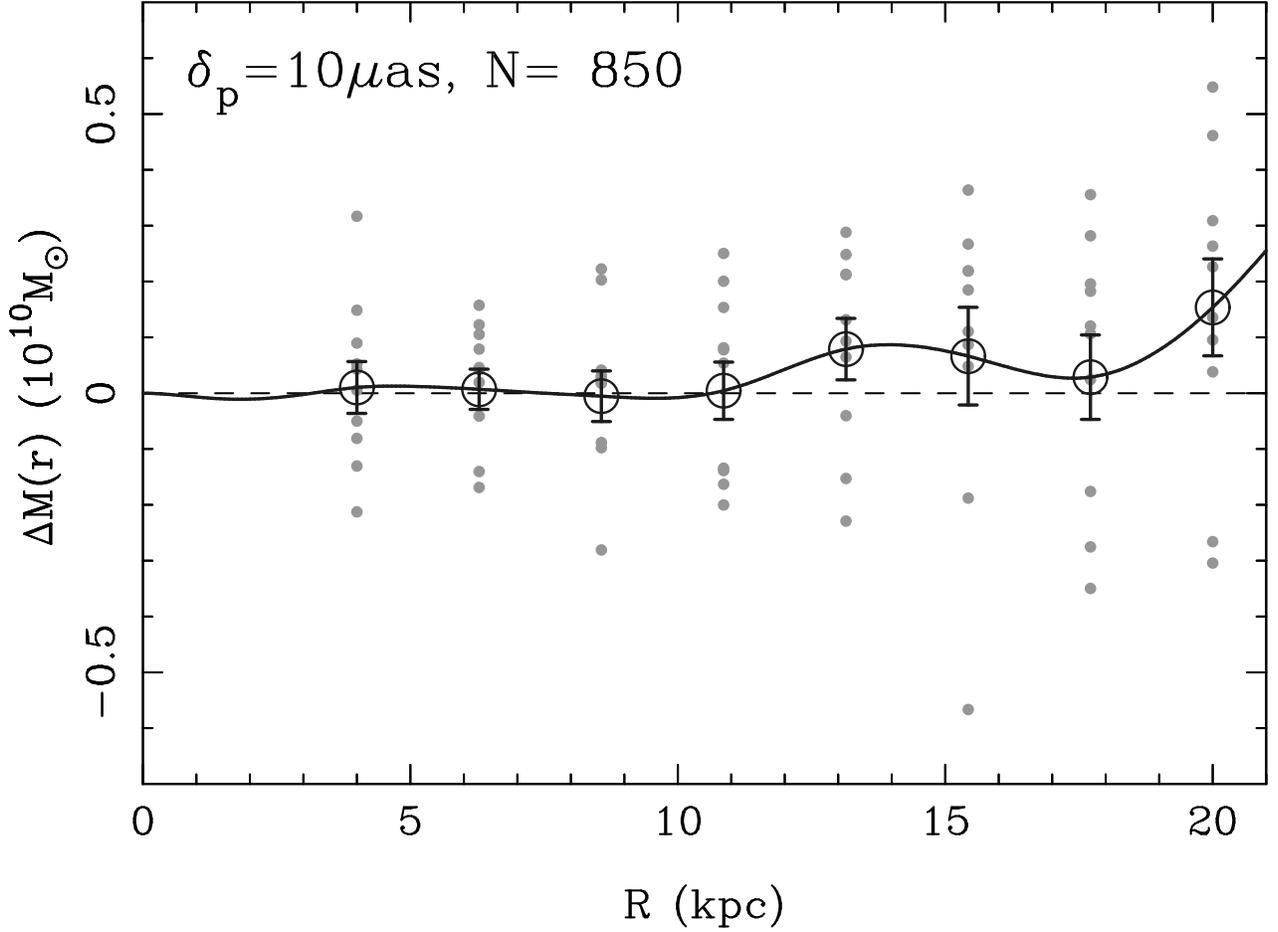}
\caption{ 
Residuals from INPUT mass, $\Delta M_i$, of 10 MCMC runs as a function of $R$ (gray dots), 
as in the middle left panel of Figure \ref{fig:spiral},
 $N=850$, $\delta_{p,{\rm phot}}=0.15p$ and $\delta_{p,{\rm tri}}=10$ $\mu$as. 
The distance to the GC, $R_0$, is one of the fitting parameters.
Here, each run uses an independent sample of simulated stars.
The open circles with error bars indicate the mean of 10 runs with 1$\sigma$ errors. 
The solid line is the interpolation of the mean $\Delta M_i$.
One can see that any systematic differences are within the error, $\sim0.1\times 10^{10} M_\sun$.
\label{fig:examerror}}
\end{figure}

In order to check for systematic biases in our methods as well as confirm the size of our error estimates,
Figure \ref{fig:examerror} 
repeats the middle left panel of Figure \ref{fig:spiral}
for 10 runs of the MCMC (gray dots) applied to 10 independent samples of simulated stars. 
The sizes of the errors  are not shown, 
but are similar to those shown in Figure \ref{fig:spiral}. 
Open circles and error bars indicate the mean $\Delta M_i$ of the 10
runs and its estimated 1$\sigma$ error, and
the solid line is the spline interpolation of the mean $\Delta M_i$.
Only two of the means lie (a little) more than 1$\sigma$ away from the
INPUT model ($\Delta M_i=0$) and thus
the scatter is consistent with the estimated statistical error
of the means estimated from the 10 runs.
In addition, the standard deviation of the OUTPUT mass parameters from their INPUT
values, $sdev= (0.20 \pm 0.02)\times 10^{10}M_\sun$
(calculated for the 10 runs over all eight points) 
is consistent with the mean of the errors estimated from the MCMC
$\bar{\delta}_M= (0.21 \pm 0.004)\times 10^{10}M_\sun$.
Overall, these comparisons validate the error estimates derived from the MCMC
(i.e., $sdev \approx \bar{\delta}_M$)
as well as the success of our method in modeling the
adopted observing strategy.

Note that the results in this section also confirm that this method 
does not require knowledge of the true spatial
distribution of our sample stars --- the algorithm we chose to pick
the sample does not enter into the analysis.
Indeed, the gray points in Figure \ref {fig:plotxy}
show that the intrinsic scatter in M-giant absolute
magnitudes means that some of our
sample lies outside of our intended survey region, $R=R_8=20$ kpc,
where the last of our mass parameters is defined.
A small systematic bias is apparent if we relax our requirement that
a star's position defined by its photometric parallax lies well within
this outer radius limit (recall, $R_{\rm max}=min(18\,{\rm kpc},R_8-0.17 d)$)
and instead include stars 
whose observed $p_{\rm phot}$ places them all the way out to  $R=20$ kpc.
Our choice of keeping our samples to within $R=18$ kpc 
keeps this bias negligible.

\begin{figure}
\epsscale{1.0}
\includegraphics[angle=-90,scale=0.65,keepaspectratio]{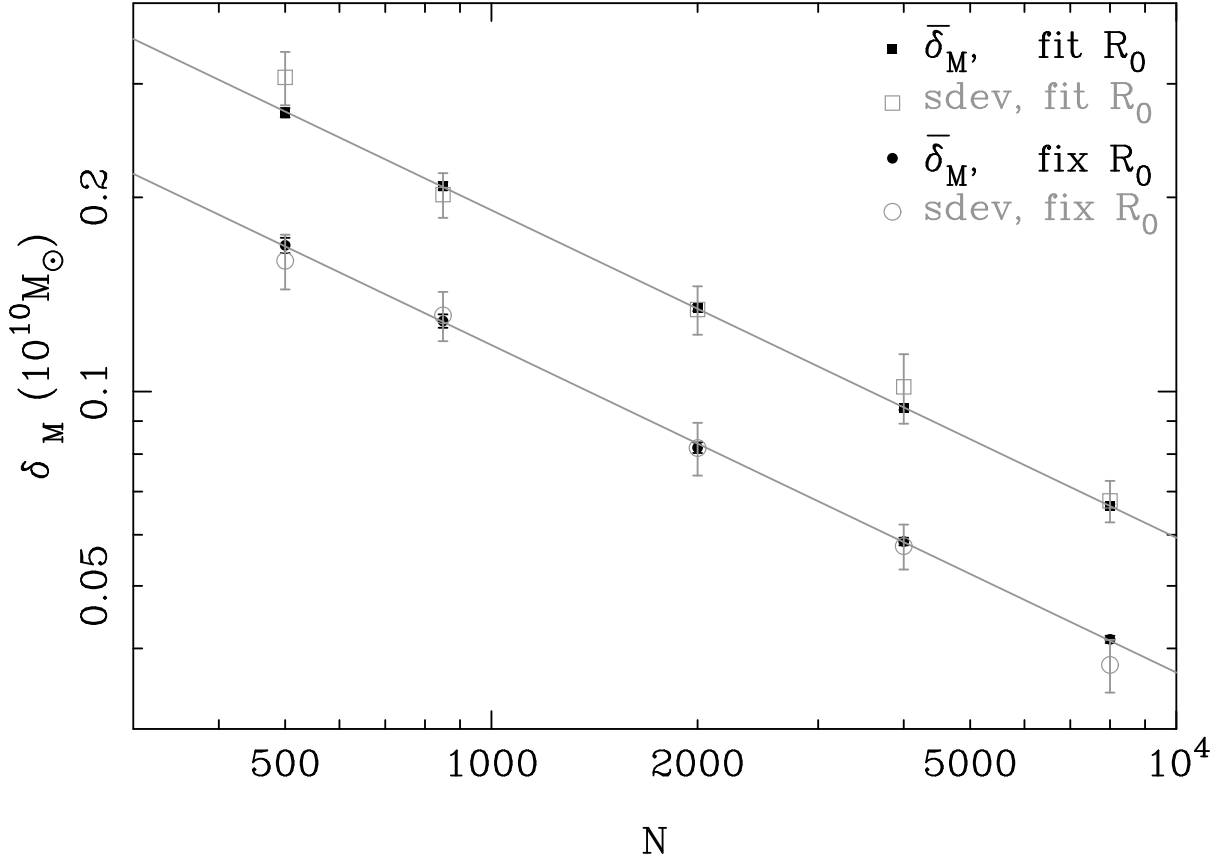}
\caption{
Mean of errors in $M_i$ calculated by the MCMC 
($\bar{\delta}_M$: filled symbols) and the standard deviation of estimates from an ensemble of 10 runs 
(sdev: open symbols with error bars) as a function of the number of stars, $N$, for
$\delta_{p,{\rm phot}}=0.15p$, $\delta_{p,{\rm tri}} =10$ $\mu$as.
The squares denote runs in which $R_0$ was fitted and the circles denote runs in
which $R_0$ was fixed at 8 kpc.
Here the mean of $\delta_M$ and $sdev$ is taken over 
$i=1$--8 and over 10 runs of MCMC.  The solid lines indicate the scaling ${\delta M /(10^{9}M_\sun}\times \sqrt{500/ N} )= 1.7$ and $2.7$, respectively.
\label{fig:checkerr-N}}
\end{figure}

\begin{figure}
\epsscale{1.0}
\includegraphics[angle=-90,scale=0.8,keepaspectratio]{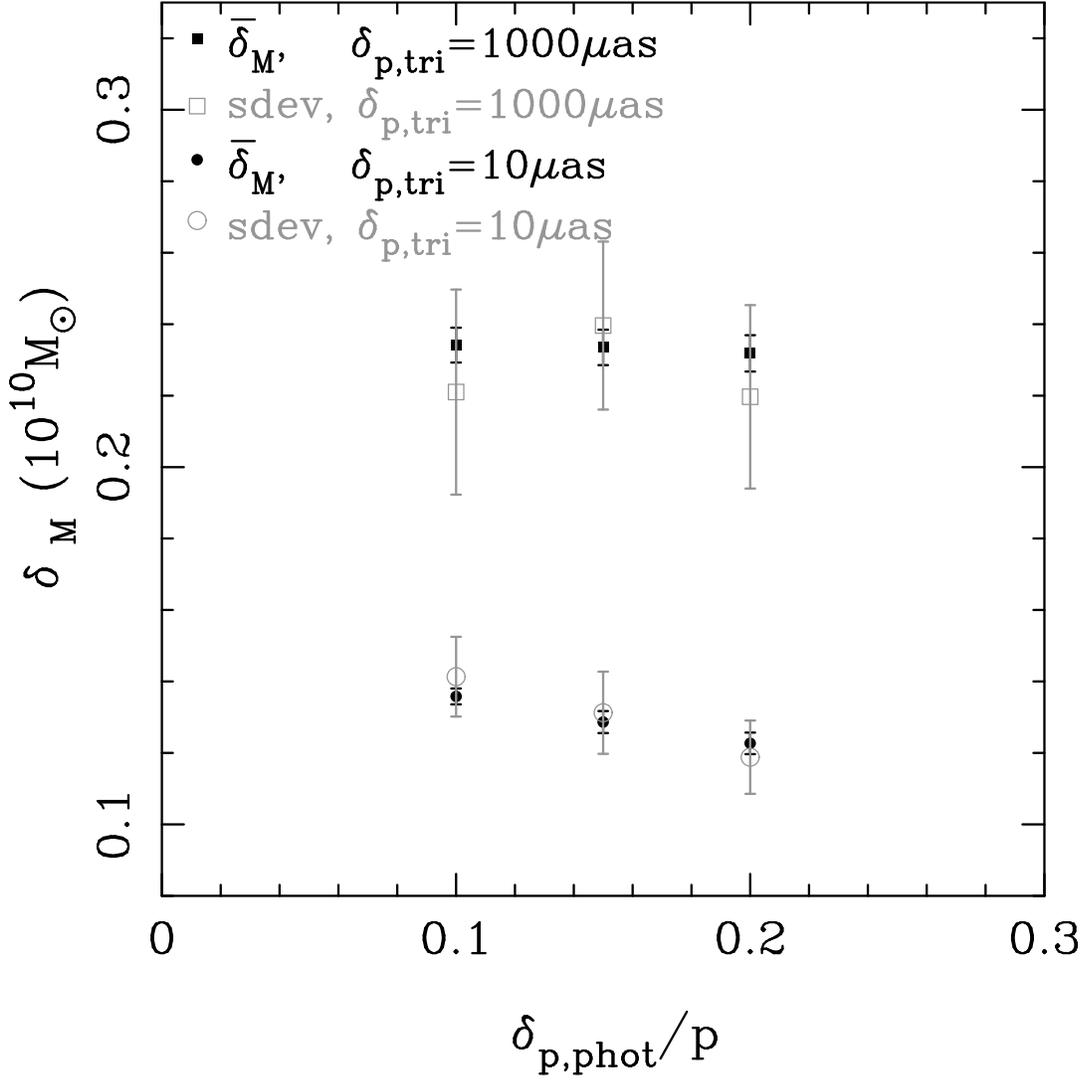}
\caption{
Mean of estimated errors in $M_i$ (filled symbols) and the standard deviation from the true
$M_i$ (open symbols) as a function of relative  photometric parallax error $\delta_{p,{\rm phot}}/p$ for $N=850$,
$\delta_{p,{\rm tri}}=10$ $\mu$as (circle) and $\delta_{p,{\rm tri}}=1000$ $\mu$as (square).
In these runs $R_0$ is fixed at 8 kpc.
Note that in some cases the error in derived quantity goes down
as the observational error goes up (see discussion in Section \ref{sec:sigdeltapphot}).
\label{fig:checkerr-pphot}}
\end{figure}

In each of the following subsections (and Figures \ref{fig:checkerr-N} - \ref{fig:checkerr-phi}),
we repeat the comparison of $sdev$ and $\bar{\delta}_M$
to check for systematic biases that may become apparent for
samples observed under different conditions.

\subsection{Dependence on $N$}
\label{sec:sigN}

Figure \ref{fig:checkerr-N} plots $sdev$ and $\bar{\delta}_M$  as a function of the 
number of stars observed $N$, with all other properties of the sample maintained at their 
standard values (see Table \ref{tbl:sdevN} for full listing of errors on all parameters).
The agreement of the
black filled and gray open symbols demonstrates both the lack of systematic biases 
in our recovery algorithm and the success of the MCMC error estimates.
The solid lines, representing the power law
\begin{equation}
\label{eqn:n}
\left({\delta_M \over 10^{9}M_\sun} \right)= 1.7 (2.7) \times \sqrt{500 \over N}
\end{equation}
for fixing (fitting) $R_0$, confirm the expected $N^{-1/2}$ scaling of errors.
The uncertainties in parameters for the case of fitting $R_0$
are increased only by order unity compared to the case of fixing $R_0$.
The uncertainty in $R_0$ as a function of $N$ is given by the formula
\begin{equation}
\label{eqn:R0-n}
\left({\delta_{R_0}  \over {\rm kpc}} \right)= 0.16 \times \sqrt{500 \over N}.
\end{equation}

\subsection{Dependence on $\delta_{p,{\rm tri}}$}
\label{sec:sigdeltatri}

\begin{figure}
\epsscale{1.0}
\includegraphics[angle=-90,scale=0.6,keepaspectratio]{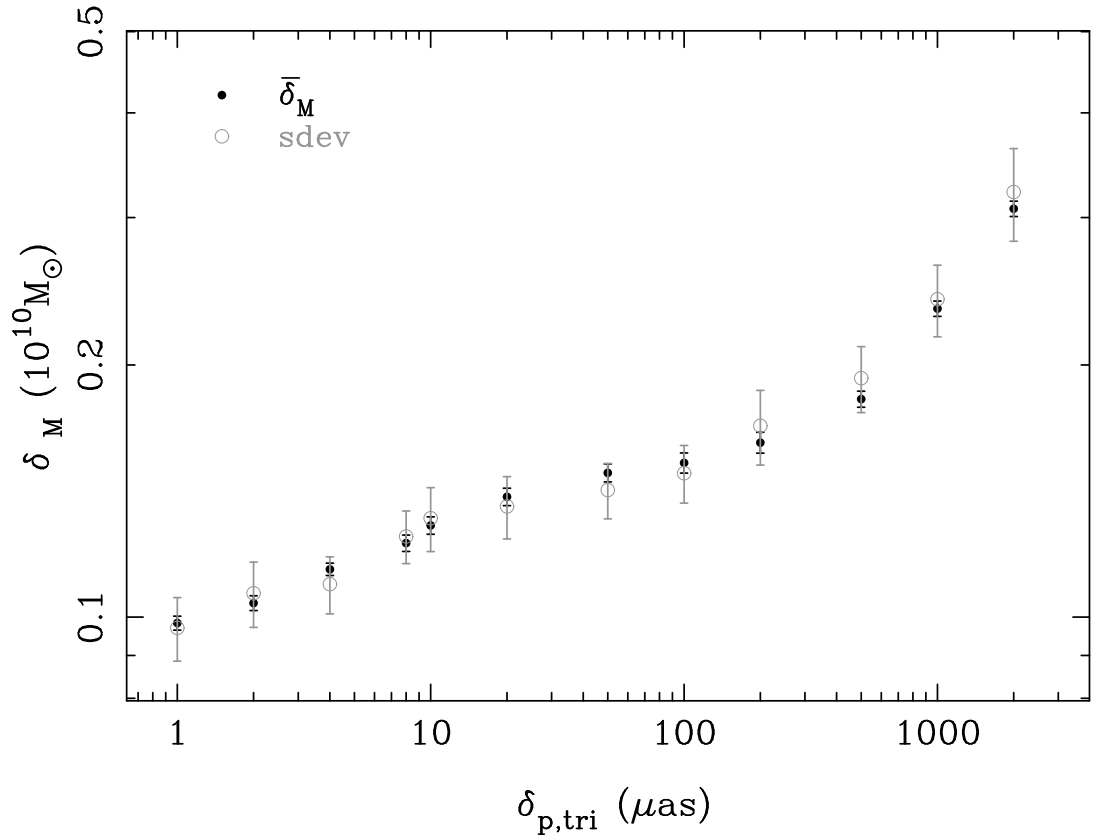}
\caption{
Mean of estimated errors in $M_i$ and the standard deviation from the true 
$M_i$ as a function of trigonometric parallax error for sample size $N=850$, 
$\delta_{p,{\rm phot}}=0.15p$ and $\delta_{p,{\rm tri}} =10$ $\mu$as. 
In these runs $R_0$ is fixed at 8 kpc.
\label{fig:checkerr-sigma}}
\end{figure}

Figure \ref{fig:checkerr-sigma} and Table \ref{tbl:sdevacc} summarize the
results of repeating the 
analysis of Section \ref{sec:standard} for MCMC runs 
based on the standard sample but with
varying trigonometric parallax errors $\delta_{p,\rm tri}$
(and related proper-motion accuracies---see Equation \ref{eq:mue}).

The figure indicates that trends of the uncertainty $\delta_M$ 
with $\delta_{p, {\rm tri}}$
can be roughly split  into three regimes.
As might be expected, $\delta_M$  increases with $\delta_{p, {\rm tri}}$ for
$\delta_{p, {\rm tri}}$ less than $\sim 10$ $\mu$as and $\delta_{p, {\rm tri}}$ 
greater than $\sim 200$ $\mu$as.
However, $\delta_M$ is almost constant for $\delta_{p, {\rm tri}}$
in the range $10-200$ $\mu$as.
This behavior can be understood by considering the importance 
of the sources of observational error in each of these three regimes.
\begin{description}
\item{For $\delta_{p, {\rm tri}}<10$ $\mu$as} the trigonometric parallax is
more accurate than the photometric parallax for most of the stars in the 
sample (those with $p<67$ $\mu$as and distances less than 15 kpc), 
while the proper-motion ($\sim6$ $\mu$as\,yr$^{-1}$, corresponding to $v_{\rm tan}<$ 1 km\,s$^{-1}$ at $<20$ kpc) 
and line-of-sight velocity accuracies (1 km\,s$^{-1}$) are much smaller than the scales 
of the velocity dispersions ($\sim 25$ km\,s$^{-1}$)
of the population that they are trying to measure. 
Hence, the uncertainty scales
with $\delta_{p, {\rm tri}}$
\item{For 10 $\mu{\rm as} < \delta_{p, {\rm tri}} < 200$ $\mu$as}, the 
photometric parallax---held constant at 15\%---provides stronger constraints on the
results than the trigonometric parallax, and the proper motion and
velocity accuracies are still too small to increase $\delta_M$.
\item{For $\delta_{p, {\rm tri}} >200$ $\mu$as}, while the photometric parallax still
provides 15\% constraints on the distances to stars, the proper-motion
error is increasing with $\delta_{p, {\rm tri}}$ 
beyond $0.21$ mas\,yr$^{-1}$, or $\sim 10$ km\,s$^{-1}$ for stars at 10 kpc.
Hence, the accuracy with which the
motions of distant stars can be determined is of the same order as their velocity dispersion
and this now
limits the accuracy with which the mass
can be measured.
\end{description}
We will discuss the implications of these trends for future surveys in Section \ref{sec:discussion}.

The overlap of the black and gray points in Figure \ref{fig:checkerr-sigma} once again confirms
the lack of systematic biases in our recovery algorithm.

\subsection{Dependence on $\delta_{p,{\rm phot}}$}
\label{sec:sigdeltapphot}

We assume 15\% photometric parallax accuracy in most of this work, i.e., 
$\delta_{p,{\rm phot}}=0.15p$.  We show the dependence of the accuracy of parameter estimates on various parallax accuracies, 
$\delta_{p,{\rm phot}}=0.1p$, $0.15p$ and $0.2p$ for $\delta_{p,{\rm tri}}=10$ and 
$1000$ $\mu$as in Figure \ref{fig:checkerr-pphot} and Table \ref{tbl:sdevpphot}.
In general, we find that the mean error on parameters {\it decreases} slightly as $\delta_{p,{\rm phot}}$ increases for 
$\delta_{p,{\rm tri}}=10$ $\mu$as, while the mean error is independent of
$\delta_{p,{\rm phot}}$ for $\delta_{p,{\rm tri}}=1000$ $\mu$as.

This counter-intuitive result---clearly in contradiction
with the  expectation that
the errors on estimates should decrease as measurements
become more accurate---can be attributed to the
different nature of the samples in each case.
Since the stars are selected using their
photometrically estimated distances, $d_{\rm phot}$
(see Section \ref{sec:observing}),
the spatial distribution of the 
true positions of stars changes 
with  $\delta_{p,{\rm phot}}$:
the larger $\delta_{p,{\rm phot}}$, the more the distribution of true
distances is determined by the parameters of the disk rather than the
parameters of the survey.

For samples that are then observed with $\delta_{p, {\rm tri}}= 10$ $\mu$as, 
the trigonometric parallax is more accurate than the photometric parallax 
for the majority of the stars ($d<15$ kpc). 
Hence, even if $\delta_{p,{\rm phot}}$ increases, 
the accuracy of the dominant distance estimates in the
analysis remains the same, while the area of the disk explored 
by the selected stars goes up and the net effect is an improvement
in the errors on the parameters.

For the same samples
observed with $\delta_{p,{\rm tri}}=1000$ $\mu$as, 
the photometric parallax
is more accurate than the trigonometric parallax for the majority of 
the stars ($>100$pc).
Now both the errors in the distance estimates and disk coverage 
increase with $\delta_{p,{\rm phot}}$, and these effects
exert competing influences on the parameter estimates.
Hence the size of uncertainties on the parameters is
largely independent of $\delta_{p,{\rm phot}}$.

The only exception to these trends in errors is for the disk scale length $h$, 
where the mean error decreases as  $\delta_{p,{\rm phot}}$ 
increases for both values of $\delta_{p,{\rm tri}}$.
This can be explained by recalling that the number density 
distribution only contributes to the likelihood function 
if the scattering of stars due to the uncertainty of $p_{\rm phot}$ is 
sufficiently large to sense the shape of the number density,
by which the likelihood function is weighted---in the idealized case of zero errors, 
the method cannot constrain $h$ at all
(see Section \ref{sec:like} and Equations \ref{eq:P}-\ref{eq:intP}).
This effect compounds rather than competes with the trend due to changing 
sample distributions as $\delta_{p,{\rm phot}}$ changes
for both values of $\delta_{p,{\rm tri}}$.

To check how the results would be affected by
systematic  errors, we ran our analysis on samples constructed
assuming photometric parallaxes that were 
10\% smaller or larger than the true parallax in
addition to the random errors. The resultant $M_i$ are systematically
overestimated and underestimated by up to 10\%, respectively. The parameters
$\Phi_{\rm a}$ and $h$ are also biased by 30\%-50\%.

\subsection{Dependence on disk coverage and knowledge of $R_0$}
\label{sec:sigmaphi}

\begin{figure}
\begin{center}
\epsscale{1.0}
\includegraphics[angle=-90,scale=0.8,keepaspectratio]{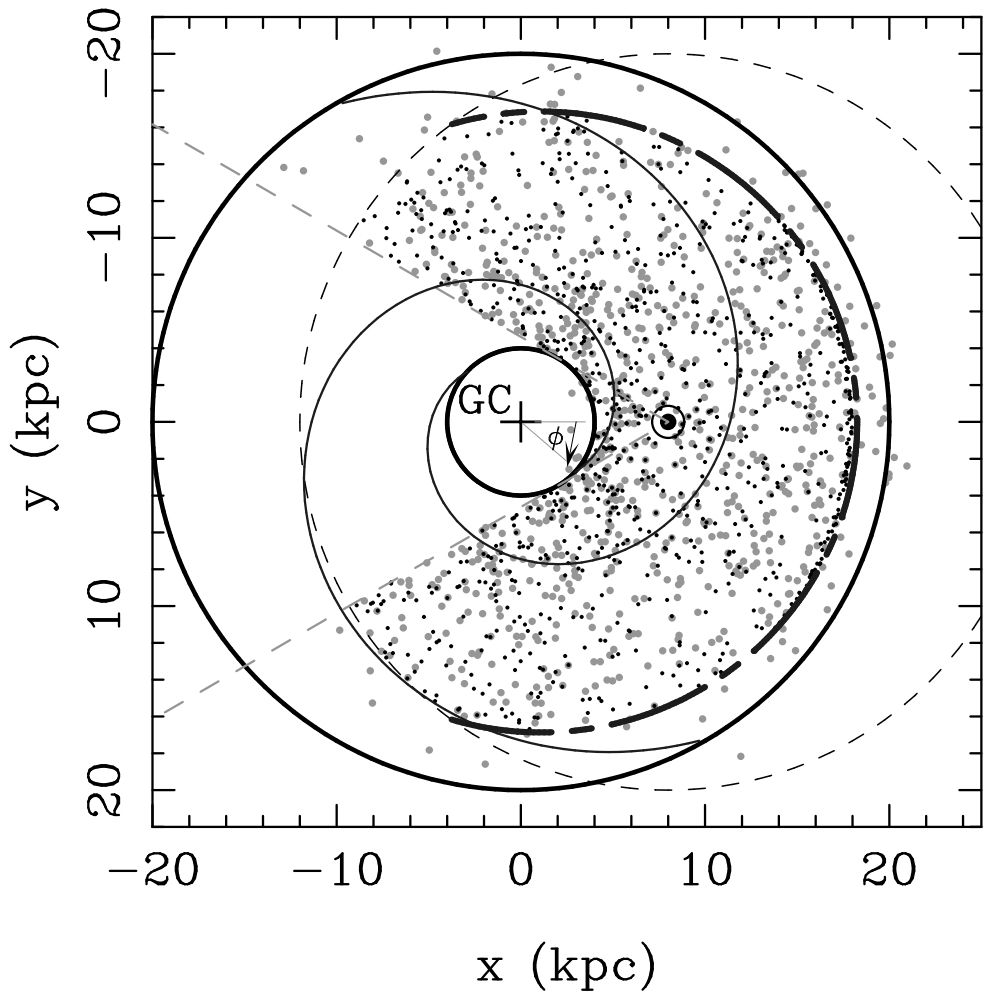}
\caption{Spatial distribution of selected stars on the
Galactic plane seen from the North Galactic Pole.
The large gray dots indicate their intrinsic positions and the small black dots represent 
their positions derived from their ``observed" photometric parallax.
The sample is chosen from their observed positions so that there are equal numbers of 
stars in the seven bins defined by the Galactocentric radii $R_i$ (Section \ref{sec:observing}).
The symbols $+$ and $\sun$
indicate the GC and the Sun. We use stars (small black dots) between
4 (inner thick circle) and 18 kpc from the GC and $<20$ kpc from the Sun (dashed circle).
We also select only stars within $20\hbox{\,kpc}$--$0.17d$ from the GC (short-long dashed line) 
to reduce the number of stars that leak outside the outermost model
radius $R_8=20$ kpc due to errors in the photometric parallax
(see details in Section \ref{sec:observing}).
Stars that lie behind the GC with $|l|<30^\circ$ (dashed lines) are not used.
The peak of the spiral arm potential is also drawn. The angle $\phi$ is defined as shown
by the arrow.
\label{fig:plotxy}}
\end{center}
\end{figure}

\begin{figure}
\epsscale{1.0}
\includegraphics[angle=-90,scale=0.6,keepaspectratio]{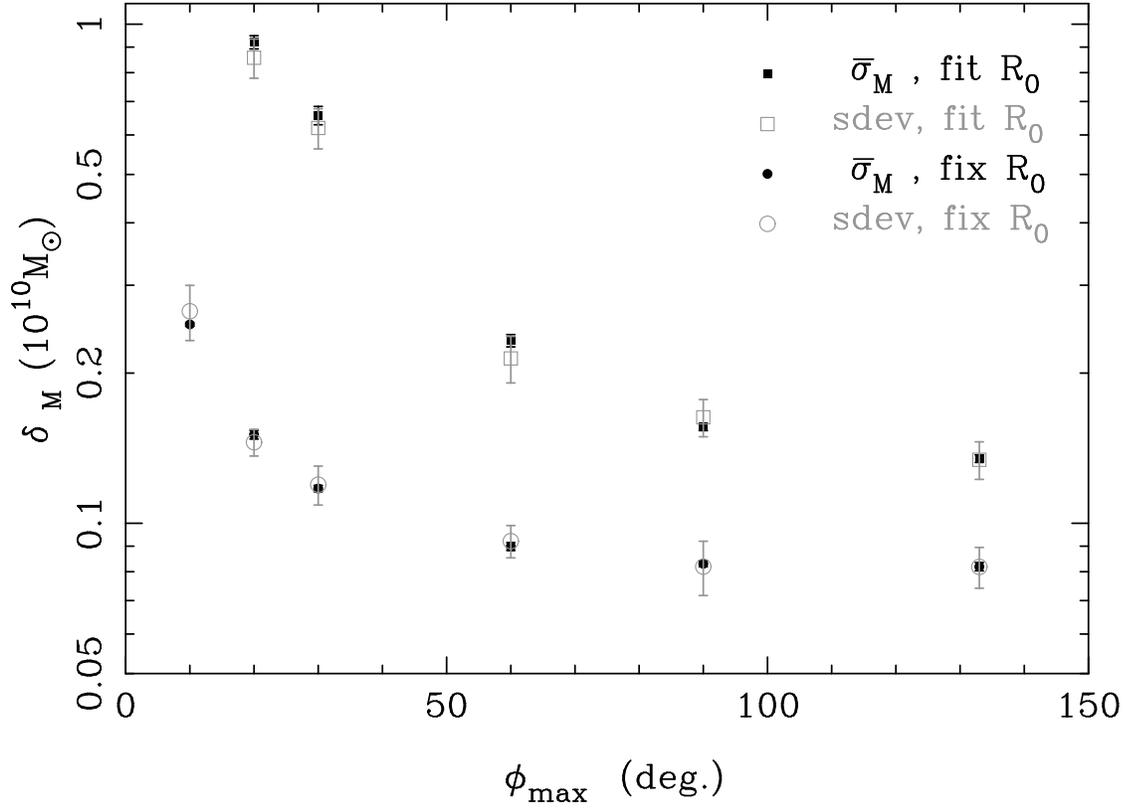}
\caption{ 
Mean estimated error in $M_i$ (filled symbols) and the standard deviation from the 
true $M_i$ (open symbols) for $N=2000$ as a function of $\phi_{\rm max}$.
The circles and squares are for the case of fixing $R_0=8$ kpc and fitting $R_0$, respectively.
The MCMC fails to converge for $\phi_{\rm max}\le10^\circ$ when $R_0$ is fitted.
Covering $\phi_{\rm max}\ge60^\circ$ is required to recover models efficiently,
especially for the case of fitting $R_0$.
\label{fig:checkerr-phi}}
\end{figure}

To check how the spatial distribution of our sample stars affects our recovery, 
we ran the MCMC for samples chosen with various values of $\phi_{\rm max}$, the maximum 
absolute azimuthal angle of stars in the sample around the GC, i.e., 
$|\phi|<\phi_{\rm max}$ (see Figure \ref{fig:plotxy}).
Here the angles $\phi$ of sample stars are estimated using $l$ and $p_{\rm phot}$.
In all previous analyses we have not restricted
$\phi$ (except that $\phi_{\rm max}\sim133^\circ$ in practice in our sample selection as seen in Figure \ref{fig:plotxy}).
The results are shown
in Table \ref{tbl:sdevphimax} (with sample sizes $N=850$ and $N=2000$)
and summarized in Figure \ref{fig:checkerr-phi} (for
$N=2000$).
As might be anticipated, the smaller the disk coverage (smaller $\phi_{\rm max}$), 
the bigger the uncertainty of parameters---it is harder to be certain of the
nonaxisymmetric features in the disk without a global view.
For our particular disk model, it was  
necessary to cover more than $\phi_{\rm max}=60^\circ$ to
recover parameters effectively.
If $\phi_{\rm max}$ was sufficiently small, the
MCMC failed to converge altogether (e.g., 
if $\phi_{\rm max}\le20^\circ$ and $\le10^\circ$ with $N=850$ and $2000$, respectively, 
when $R_0$ is fitted).

Figures \ref{fig:checkerr-N} and \ref{fig:checkerr-phi} and 
Tables \ref{tbl:sdevN} and \ref{tbl:sdevphimax} illustrate
the results of experiments both for the case of fitting the distance $R_0$ to the GC
and the case of fixing $R_0=8$ kpc assuming an accurate assessment of $R_0$
has been made from other sources \citep[e.g.,][]{eis03}.
If the disk is surveyed globally (i.e., for large $\phi_{\rm max}$),
$R_0$ can be recovered with a few percent accuracy
(see Table \ref{tbl:sdevN} and \ref{tbl:sdevphimax})
using the adopted samples, and the uncertainties in other parameters 
are increased only by order unity compared to the case of fixing $R_0$.
However, for $\phi_{\rm max}<60^\circ$ the uncertainty in $R_0$ and all
other parameters increases even more dramatically with $\phi_{\rm max}$ compared
to the examples where $R_0$ was fixed.

\section{Discussion: implications for near-future surveys}
\label{sec:discussion}

\subsection{Astrometric surveys}

\subsubsection{NASA's Space Interferometry Mission (SIM) Lite}
\label{sec:sim}

\begin{figure}
\epsscale{1.0}
\includegraphics[angle=-90,scale=0.65,keepaspectratio]{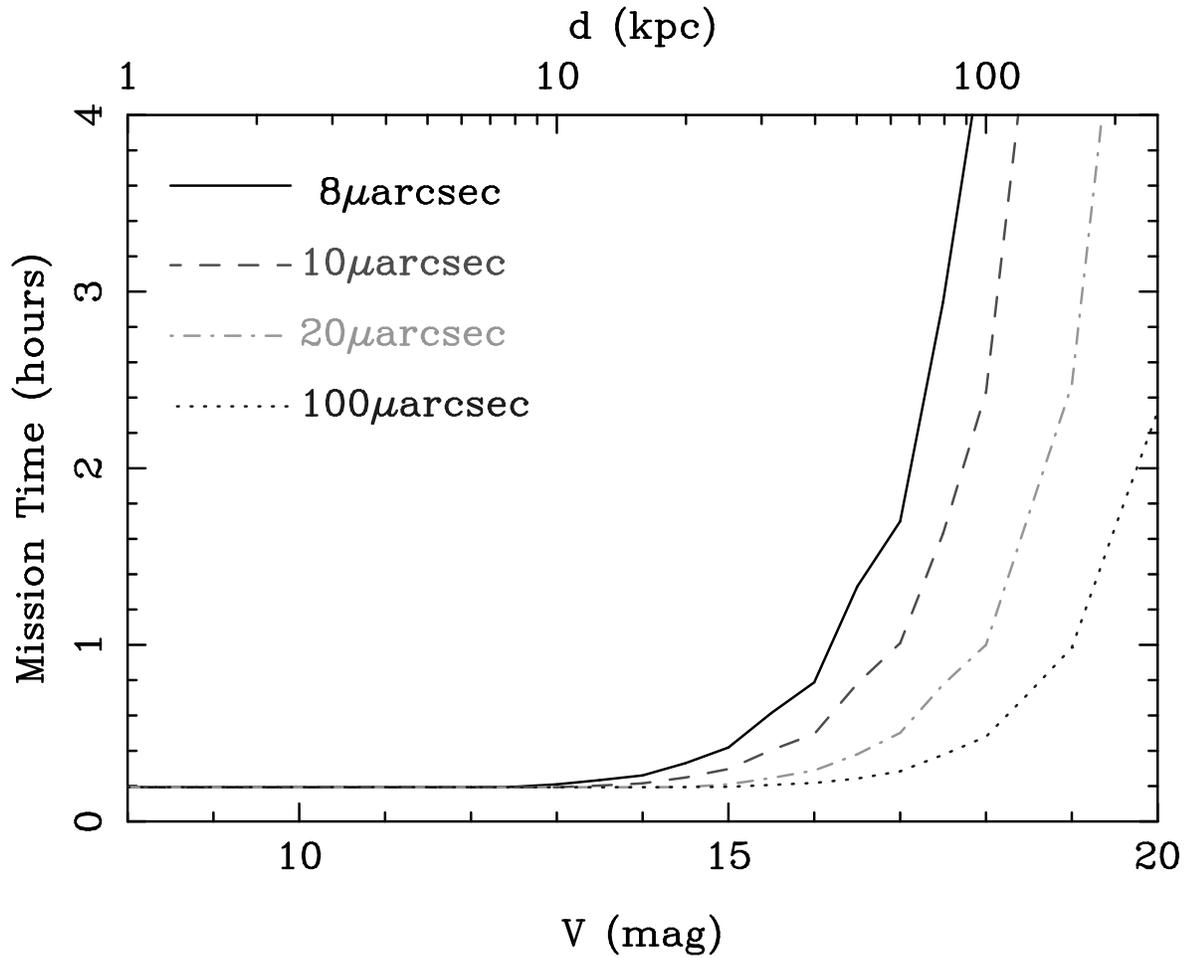}
\caption{SIM Lite's required mission time to measure the parallax of a star with
8, 10, 20 and 100 $\mu$as accuracy as a function of V-band magnitude of the star.
The top horizontal axis indicates the distance to the target with an absolute magnitude 
of $M_{V}=-2$ mag (e.g., M giants).
\label{fig:missiontime}}
\end{figure}

SIM Lite is a planned astrometric satellite using an optical 
interferometer, which will yield parallax errors as small as 4 $\mu$as in wide angle mode.
SIM Lite observes a parallax and a proper motion by repeatedly pointing at a target star 
and integrating as long as is necessary to get the required accuracy. 
For example, Figure \ref{fig:missiontime} shows the expected mission time
for stars requiring accuracies $\delta_{p,{\rm tri}}= 8$, 10, 20 and 100 $\mu$as 
observed in wide angle mode up to the limiting magnitude $V=20$.
\footnotemark\footnotetext{\tt http://mscws4.ipac.caltech.edu/simtools/portal/login/normal/1?}

Observers are allocated a specific amount of time to design their own experiments.
Hence, for any disk study it is vital to find the optimal number, spatial coverage and required
trigonometric accuracies of target stars to recover the Galactic mass distribution most effectively.
Figure \ref{fig:checkerr-sigma} and Table \ref{tbl:sdevacc} suggest that
the uncertainties in our mass estimates are only weakly dependent on $\delta_{p,\rm tri}$ in
the 10-100 $\mu$as regime. On the other hand, Figure \ref{fig:checkerr-N} and Table \ref{tbl:sdevN} 
suggests that the uncertainty scales as $\sqrt{1/N}$ 
(as given in Equations \ref{eqn:n} and \ref{eqn:R0-n}).
These results imply that our best strategy for studies of the mass
distribution in the Galactic disk using SIM Lite (assuming that the astrometric errors are indeed statistical, not systematic) is
to choose as large a sample as possible within the given observing time
rather than using the time to get the best astrometric accuracies for a
smaller number of targets.

\begin{figure}
\epsscale{1.0}
\includegraphics[angle=-90,scale=0.5,keepaspectratio]{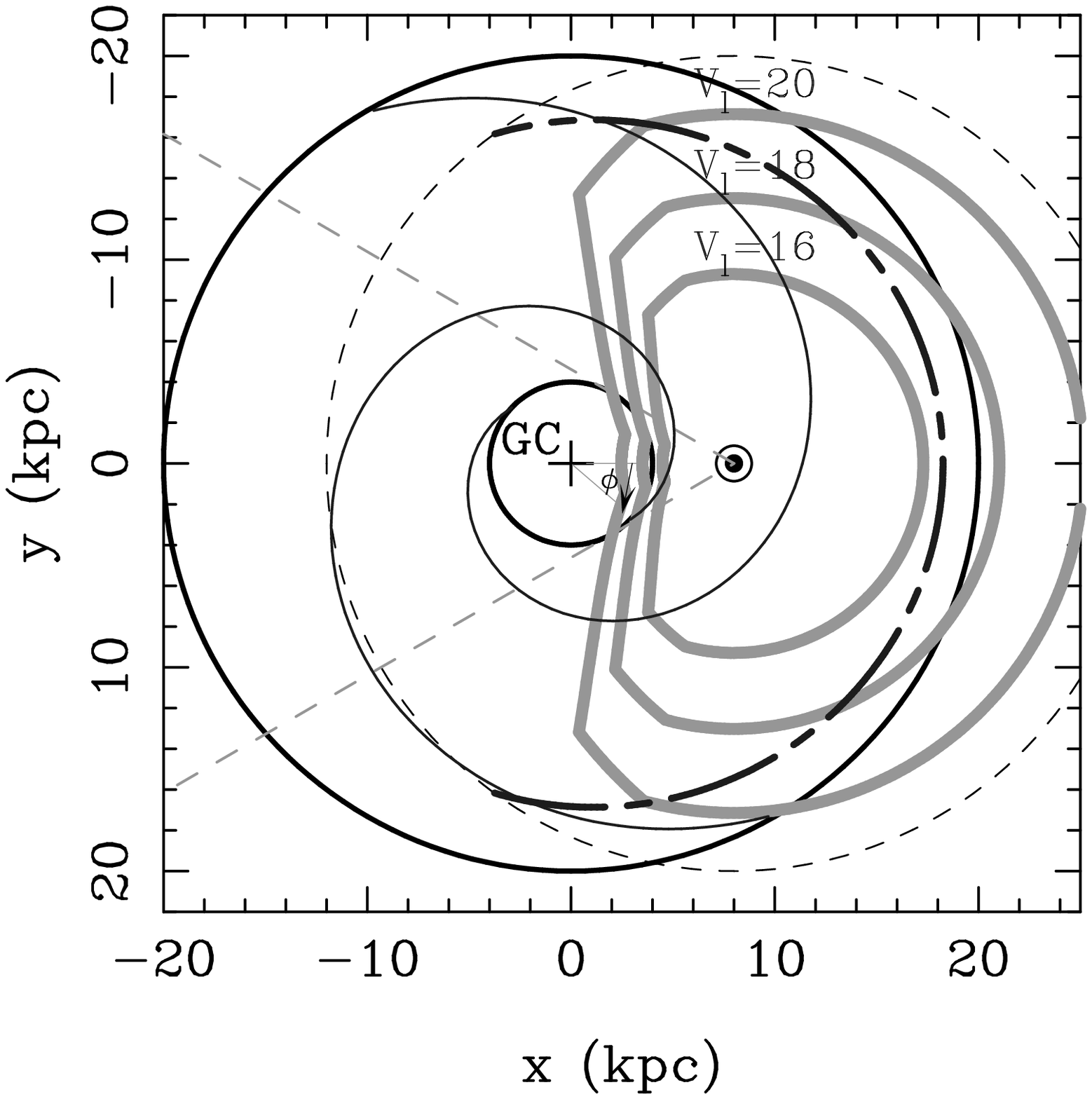}
\includegraphics[angle=-90,scale=0.5,keepaspectratio]{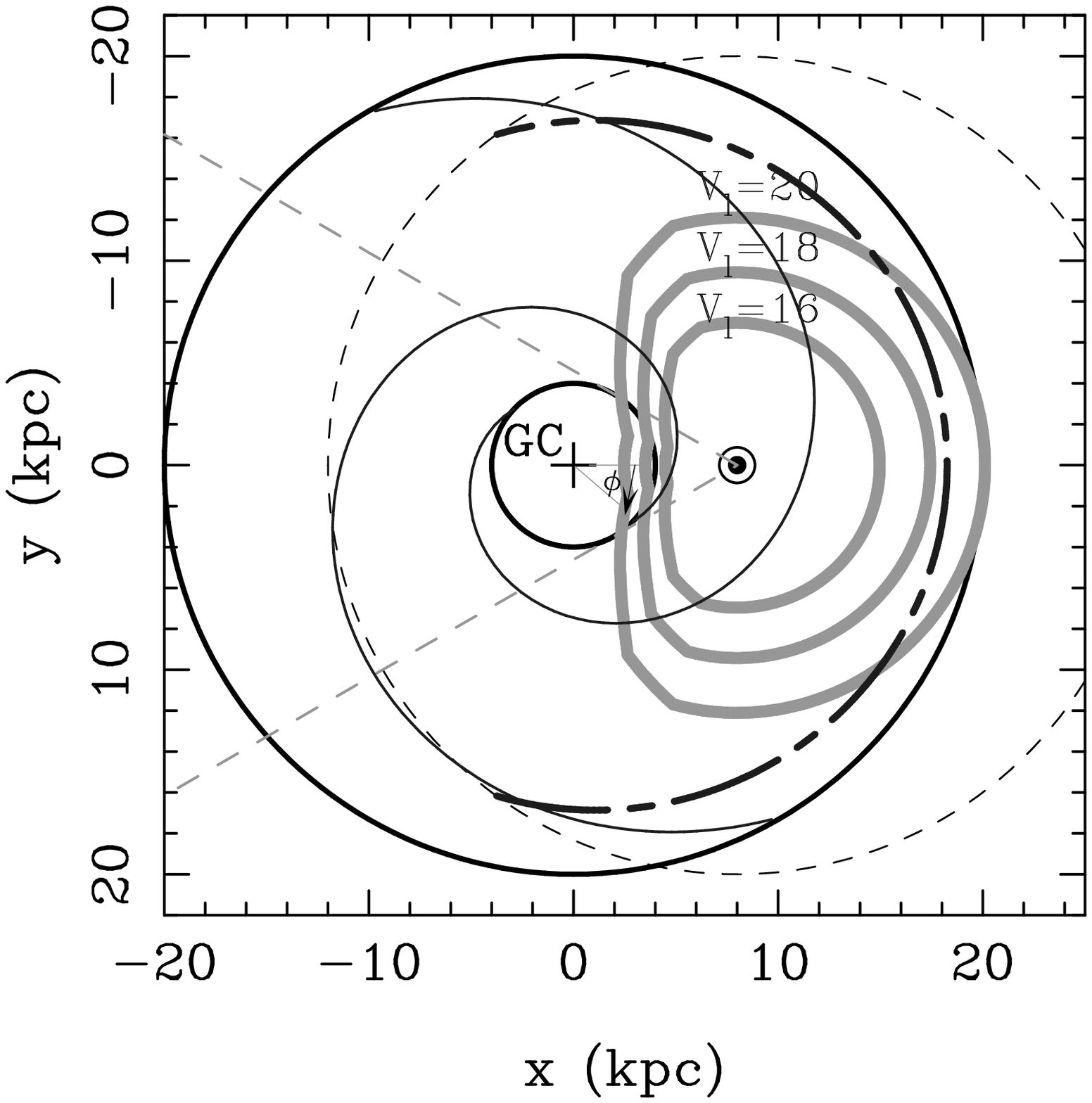}
\caption{Same as Figure \ref{fig:plotxy}, but showing the maximum survey region with 
the $V$-band limiting magnitude $V_{\rm l}=16$, 18 and 20 mag (the thick gray lines) 
for the case that we observe M-giants with an 
absolute magnitude of $M_V=-2$ through the low extinction windows with 
$A_{V,\rm win}/{A_{V,\rm typ}}$ =0.5 (left panel) and 0.8 (right panel)  (see details in Section \ref{sec:sim}).
\label{fig:plotxy-AV}}
\end{figure}

In principle, the number of stars observed by SIM Lite could be maximized by looking 
only at objects brighter than some limiting magnitude, $V_{\rm l}$. 
In order to make a realistic assessment of the extent to which such a magnitude limit 
would compromise our spatial coverage we need to also account for extinction in the 
disk plane. G. Zasowski, private communication, found typical $V$-band extinctions in the Galactic disk to be:
$A_V \sim (1.5-8)\times d$\,kpc$^{-1}$ at $|l| <  15^{\circ}$,
$A_V \sim 0.8\times d$\,kpc$^{-1}$ at $|l| = 60^{\circ}$ and
$A_V \sim 0.68\times d$\,kpc$^{-1}$ at $|l| = 90^{\circ}$ and $180^{\circ}$.
With our small sample size ($N \sim 1000$), we assume we can restrict our attention to stars that can be observed in low-extinction windows.
Hence we adopt the lowest value, $A_V \sim 1.5\times d$\,kpc$^{-1}$ for $|l| < 15^{\circ}$.
For $|l| > 15^{\circ}$, the typical values are linearly interpolated in $l$, 
and a fixed fraction $A_{V,\rm win}/{A_{V,\rm typ}}$ of the estimated 
extinction is used.
Figure \ref{fig:plotxy-AV} illustrates the spatial coverage attainable for M-giants brighter than
$V_{\rm l}=16/18/20$ and with $A_{V,\rm win}/{A_{V,\rm typ}}=0.5$ (left-hand panel) and 0.8 (right-hand panel).
While the extinction keeps us from seeing entirely across the Galactic disk in all cases, 
the requirement that $\phi_{\rm max} > 60^\circ$ in order to account for disk asymmetries (see Section \ref{sec:sigmaphi}) is still met, at least at small radii.
 
Table \ref{tbl:NobsSIM} shows the number of M-giant stars that could be observed by SIM Lite 
within an allocation of 240 hr (motivated by the typical sizes of Key Project proposals for 
SIM PlanetQuest, the predecessor mission to the SIM Lite mission, which is
similar, but based on a modified instrument architecture)
for various $V_{\rm l}$ and $A_{V,\rm win}/{A_{V,\rm typ}}$, adopting the strategy outlined in previous sections of sampling equal numbers of stars in the 7 radial bins between the $R_i$ in our model  (see details in Section \ref{sec:observing}). The blank entries in the table correspond to cases where one or more of the radial bins contained no stars. 
Indeed,  while the brighter limiting magnitudes do allow more stars to be observed (i.e., larger $N$), the outer radial bins in these cases either contain no stars, or are only populated over a small range in $\phi$. 

Overall, we find that the optimal sample has $N\sim 1000$, $V_{\rm l}=18$ and $\delta_{p, \rm tri}=100 \mu$as. Table \ref{tbl:deltaMSIM} illustrates this by showing the uncertainty $\bar{\delta}_M$ for samples with $\delta_{p, \rm tri}=100 \mu$as and various $V_{\rm l}$.
Even for the worst case of $A_{V,\rm win}/{A_{V,\rm typ}}=1.0$, the masses can be constrained
with an uncertainty of $\bar{\delta}_M=0.31\times 10^{10} M_\sun$, which corresponds 
about $2\%-4$\% accuracy at $R=6-20$\,kpc.

\subsubsection{ESA's GAIA satellite}

Unlike SIM Lite, GAIA is an all-sky survey satellite, where each star brighter than $V=20$
is observed for the same amount of time.
Concentrating again on the M-giants, this limiting magnitude means that the GAIA 
catalog will contain the stars in the disk as shown in Figure \ref{fig:plotxy-AV},
although they will not be distributed uniformly in Galactocentric radius (because of the disk's
intrinsic density gradient)
and the variation of extinction along different lines
of sight means that the depth of the sample will not be constant.
In addition, the fainter stars will have less accurate astrometric measurements
than the brighter ones ($\delta p_{\rm tri}=21$ $\mu$as for $V=15$ and 275 $\mu$as
for $V=20$, with the corresponding proper-motion accuracies of 11 $\mu$as\,yr$^{-1}$ and 
145 $\mu$as\,yr$^{-1}$
\footnotemark\footnotetext{\tt http://www.rssd.esa.int/index.php?project=GAIA\&page=Info\_sheets\_overview}
).

Figure \ref{fig:checkerr-sigma} suggests that the target proper motion
accuracies (better than $\sim 100$ $\mu$as\,yr$^{-1}$) are at the appropriate 
level to accurately assess the kinematical properties
of disk stars when coupled with a photometric distance estimate.
Nor do we expect the non-uniformity in the sample
to introduce biases in results as
our method is independent of the spatial distribution of the target stars.
The gradient in the disk density coupled with the magnitude-limited nature of the
survey means that the outer parts of the disk will be much more sparsely sampled and 
with larger error bars on the observations than the inner parts, so uncertainties in 
the mass estimates at large Galactocentric radii will increase correspondingly.
In addition, since GAIA, like SIM, works in the optical, 
significant coverage beyond the Galactic center
may be impossible due to extinction effects.
However, these uncertainties can perhaps be offset by
the sheer number of stars in the catalog.
For example, the 2MASS catalog contains millions of M-giant candidates (i.e., in the color range $0.95<J-K<1.2$) 
brighter than $K=14$ (corresponding to $V=18$)
within 10 deg of the Galactic plane.
While there will be significant contribution by bulge stars
in this sample for Galactic longitudes $|l|<10^\circ$, they could be accounted for by adjusting the model to include
this extra component.
This argument assumes that the
astrometric errors in GAIA are mainly statistical; given samples of
this size, the biggest uncertainty with GAIA may be how
large $N$ can be before systematic errors begin to dominate the error budget.
Combining the GAIA analysis with results from surveys that
can fully cover the extent of the disk (e.g., SIM Lite; see Section \ref{sec:sim}) 
with well-characterized systematics 
(e.g., VERA; see Section \ref{sec:vera})  should yield a global picture of the mass distribution within
the Galactic disk.

\subsubsection{Radio VLBI arrays}
\label{sec:vera}

VERA \citep{hon00}, VLBA \citep{reid08,hachisuka09}, and EVN \citep{rygl08} are radio VLBI arrays 
that are conducting  $\sim10\mu$as astrometric
observations for water and/or methanol masers. 
Compared to our M-giant sample, 
water masers have the advantage as targets that most of them lie in
star-forming regions very close to the Galactic plane.
Hence they have rather lower velocity dispersion
and fewer sources may be needed
 to accurately trace the rotation curve (and mass distribution).
A possible disadvantage is that star-forming
regions and therefore the maser targets are generated in spiral arms
and thus tend to be found in a narrow range of azimuthal phases
relative to the arms. In this case there may be larger uncertainties
in and strong covariances between some of the parameters.
Another disadvantage of the masers formed in young
star-forming regions is that these targets may not yet be dynamically
homogenized to the disk, and so their motions may reflect the particular
dynamics of their particular star-forming region, including such things
as a vertex deviation, or other peculiar motions.

The most exciting property of these samples is that they are already being produced.
For example, 18 masers with parallax errors $6$-$80\, \mu$as by VLBA and VERA are
already in the literature \citep{reid09}, and VERA is expected to observe approximately
 1,000 such sources over the next 10 years
This sample will have observational error
properties very similar to those assumed for our standard sample
(Section \ref{sec:standard}), except that photometric parallaxes are not available.
Unlike our proposed M-giant survey, the distribution of the sources 
(mostly star-forming regions and some Mira variables) cannot be
chosen arbitrarily, so the disk will not be uniformly sampled--in particular, 
water masers are more rare at large Galactocentric radii ($>10$ kpc) and
it may be hard to achieve full disk coverage.
However, our results (e.g., Figure \ref{fig:checkerr-sigma}) 
suggest that a sample of this size and level of accuracy could provide 
strong constraints on the mass distribution in the inner Galaxy.

In addition to representing a significant step forward in 
measuring the Milky Way's rotation curve, 
these observations can serve as a vital
cross check for future satellite surveys that
rely on more sophisticated technology and hence may be
more prone to unanticipated systematic biases.
Furthermore, these radio VLBI observations do not suffer from dust extinction
that the optical astrometry satellites GAIA and SIM Lite will.

\subsection{Spectroscopic surveys}

High-resolution multi-object spectrographs are also currently 
under development.
For example, 
the Apache Point Observatory Galactic Evolution Experiment (APOGEE---part of 
the Sloan Digital Sky Survey III project)
will carry out a massive radial velocity
survey starting in 2011.
APOGEE  will use a 300 fiber 
near-infrared (H-band) spectrograph with $R=22,500-25,000$ to measure 
radial velocities to better than $0.5$ km\,$s^{-1}$ for more than $10^5$ stars 
predominantly across the bulge and disk.  
The survey will take advantage of the low reddening in the near-infrared 
to reach stars throughout the
Galactic disk and exploit the high resolution to 
make accurate estimates of spectroscopic parallaxes.
Our results from Section \ref{sec:sigdeltatri} indicate that coupling these
derived spectroscopic parallaxes
with proper-motion measurements accurate to only $\sim 1$ mas\,yr$^{-1}$
could provide strong constraints on the mass distribution,
especially given the large sample size.
The Large Synoptic Survey Telescope could provide these proper motions for
the third and fourth Galactic quadrants, down to a limiting magnitude of $r=24$, 
allowing stars to be surveyed all the way across the Galactic disk. 
Partial coverage of the first and second quadrants can be achieved with similar accuracies (though brighter limiting magnitude) by combining the Sloan Digital Sky Survey with the Palomar Observatory Sky Survey \citep{munn04}.
This approach would offer an alternative, independent assessment of the mass distribution to complement the astrometric measurements described above.

\section{Summary and Conclusions}
\label{sec:conclusion}

In this paper, we examined how accurately we might be able to recover the 
mass distribution in the Galaxy (or more precisely, the gravitational force 
field in the Galactic disk) using current and 
near-future astrometric, photometric, and spectroscopic
surveys of disk stars.
We simulated observations of stars drawn from a simple model for the phase-space structure in 
an equilibrium, nonaxisymmetric disk
and used a Markov Chain Monte Carlo approach to attempt to recover the model's parameters.
Our formulation of this method relied on finding the parameter set of the model that
maximizes the probability of stars in a survey having
their observed trigonometric parallax, proper motion, and line-of-sight velocity 
given their measured photometric parallax.
Hence, it is immune to biases in sample selection.
Indeed, with a correct representation of the error distributions for each observable 
and of the non-axisymmetric components of the mass distribution,
no systematic errors were evident in our approach for samples of 100s-1000s of stars
combining observed trigonometric
parallaxes with errors in the range 1 $\mu$as to 2 mas (corresponding to 
proper motion errors assumed to be in the range 0.9 $\mu$as\,yr$^{-1}$ to 1.2 mas\,yr$^{-1}$), 
photometric parallaxes known at the 10\%-20\% level and
line-of-sight velocity measurements accurate to 1 km\,s$^{-1}$.

The presence of non-axisymmetric features in the disk means that
a precise mapping of the Galactic mass distribution will require 
a survey on global scales.
Even in our simplified case with a single two-armed spiral pattern, restricting 
our survey to a limited range of Galactic longitude
significantly reduced the accuracy of our results.

However, given such a global disk survey, we found that we could recover
the mass profile in the range 4-20 kpc with a few percent accuracies
using a variety of approaches.
If $\mu$as trigonometric parallaxes are available 
(with associated proper-motion measurements and 1 km\,s$^{-1}$ line-of-sight velocities),
then this accuracy is feasible with a survey as small as a few hundred stars.
Once trigonometric parallax errors exceed 10 $\mu$as, the same accuracy can be
achieved by supplementing the trigonometric parallaxes with
photometric parallaxes accurate to 10\%-20\% and 
adopting a sample of thousands of stars, so long as proper-motion errors
remain below a level of few hundred $\mu$as\,yr$^{-1}$.
If proper-motion errors are of order a few mas\,yr$^{-1}$, then larger samples are
needed in compensation.

We also found we could measure the mass distribution even in
the absence of an accurate assessment of the distance to the 
Galactic center, $R_0$.
Including $R_0$ as a free parameter did increase our uncertainties 
by a factor of 2, but also allowed us to measure this distance 
with comparable accuracy  to the mass distribution itself (i.e., a few percent
for the samples discussed above).

We conclude that, whether one or all of the 
future surveys (e.g., SIM Lite, GAIA, VERA and APOGEE) are completed,
a significant step forward in our understanding of the Galactic mass distribution
 (i.e., an assessment of the force field in the Galactic disk at the 1\% level)
is on the horizon,
as well as detailed insights into disk dynamics.

These conclusions are based on the assumption that the deviations from a
smooth axisymmetric model of the gravitational field in the disk can
be modeled as a grand-design spiral pattern with a specified form
(logarithmic spiral) and a well-defined pattern speed, without
resonances within the disk. Further work is required to understand
how relaxing these assumptions would affect the accuracy of
astrometric disk surveys. 
Natural directions for future works are including
vertical motions of sample stars as well as a bulge component,
and more general models of spiral structure.

\acknowledgments 
We acknowledge R. Patterson and J. Carlin for useful assistance with calculations 
regarding SIM and 2MASS.  We thank L. S. Eyer for providing
important insights into GAIA.
We are grateful to M. Honma for the comments about VERA. 
We thank G. Zasowski for help with the dust extinction estimates used here.
This work is supported by  NASA/JPL contract 1228235 for the 
``Taking Measure of the Milky Way'' key project of the Space Interferometry Mission.

\clearpage
\begin{deluxetable}{crrrrrr}
\tablecaption{Parameters with INPUT values and an example OUTPUT.  \label{tbl:param}}
\tablewidth{0pt}
\tablehead{
\colhead{Parameter} & $R_{i}({\rm kpc)}$ & \colhead{INPUT} & \colhead{OUTPUT} & \colhead{$1\sigma$ Error} \\
}
\startdata
             $M_{1} (10^{10}M_\sun)$ &   4.00 &     2.37 &     2.24  &     0.15 \\
             $M_{2} (10^{10}M_\sun)$ &   6.29 &     4.94 &     4.90  &     0.10 \\
             $M_{3} (10^{10}M_\sun)$ &   8.57 &     7.86 &     7.77  &     0.12 \\
             $M_{4} (10^{10}M_\sun)$ &  10.86 &    10.92 &    10.76  &     0.16 \\
             $M_{5} (10^{10}M_\sun)$ &  13.14 &    13.99 &    13.83  &     0.17 \\
             $M_{6} (10^{10}M_\sun)$ &  15.43 &    16.99 &    17.10  &     0.27 \\
             $M_{7} (10^{10}M_\sun)$ &  17.71 &    19.90 &    19.72  &     0.30 \\
             $M_{8} (10^{10}M_\sun)$ &  20.00 &    22.68 &    22.77  &     0.33 \\
 $\sigma_{R,\sun} {\rm (km\,s^{-1})}$ &    --- &    25.00 &    24.64  &     0.64 \\
                    $\phi_{0}$ (rad) &    --- &     1.83 &     1.85  &     0.07 \\
                           $c$ (rad) &    --- &     7.46 &     7.51  &     0.17 \\
 $\Omega_p$ (km\,s$^{-1}$\,kpc$^{-1}$) &    --- &     1.40 &     1.12  &     0.37 \\
     $\Phi_{a} {\rm (km\,s^{-1})^2}$ &    --- &   200.0  &    208.3  &    32.8 \\
                  $h_{\sigma}$ (kpc) &    --- &     3.00 &     3.14  &     0.10 \\
                           $h$ (kpc) &    --- &     3.00 &     3.36  &     0.35 \\
                         $R_0$ (kpc) &    --- &     8.00 &     7.97  &     0.12 \\

\enddata
\tablecomments{
Recovered OUTPUT parameters and 1$\sigma$ errors for a survey with $\delta_{p,{\rm phot}}=0.15p$
and $\delta_{p,{\rm tri}}=10$ $\mu$as and $N=850$ stars.  $M_i$ is the mass within $R_i$. 
}
\end{deluxetable}

\begin{deluxetable}{rrrrrrrrrr}
\tabletypesize{\scriptsize}
\tablecaption{Mean errors of parameters Versus number of stars $N$.
\label{tbl:sdevN}}
\tablewidth{0pt}
\tablehead{
$N$ & $\bar{M}$ & $\sigma_{R,\sun}$ & $\phi_0$ & $c$ &  $\Omega_p$ & $\Phi_{a}$ & $h_\sigma$ & $h$ & $R_0$\\}
\startdata
 500 &  0.17 &  0.79 &  0.08 &  0.23 &  0.54 &  48.2 &  0.12 &  0.44 & ---\\
 850 &  0.13 &  0.60 &  0.06 &  0.17 &  0.41 &  36.7 &  0.09 &  0.29 & ---\\
 2000 &  0.08 &  0.39 &  0.04 &  0.10 &  0.30 &  27.5 &  0.06 &  0.19 & ---\\
 4000 &  0.06 &  0.28 &  0.02 &  0.07 &  0.20 &  19.1 &  0.04 &  0.12 & ---\\
 8000 &  0.04 &  0.20 &  0.02 &  0.05 &  0.13 &  12.4 &  0.03 &  0.09 & ---\\
\hline
 500 &  0.27 &  0.88 &  0.09 &  0.22 &  0.52 &  44.8 &  0.12 &  0.43 &  0.16\\
 850 &  0.21 &  0.66 &  0.07 &  0.17 &  0.42 &  37.5 &  0.09 &  0.32 &  0.12\\
 2000 &  0.13 &  0.42 &  0.05 &  0.11 &  0.30 &  27.2 &  0.06 &  0.19 &  0.08\\
 4000 &  0.09 &  0.30 &  0.03 &  0.07 &  0.21 &  19.1 &  0.04 &  0.13 &  0.05\\
 8000 &  0.07 &  0.21 &  0.02 &  0.05 &  0.13 &  11.7 &  0.03 &  0.09 &  0.04\\
\enddata
\tablecomments{
The means of the estimated error of parameters are calculated from 10 runs 
of MCMC for the case of $\delta_{p,{\rm tri}}=10$ $\mu$as, either fixing $R_0=8$ kpc (shown as ---)
or fitting $R_0$.
Units are same as in Table \ref{tbl:param}. For $\bar{M}$, the mean is 
also taken over $i=$1--8.  
See the notes for Table \ref{tbl:param}.
}
\end{deluxetable}

\begin{deluxetable}{rrrrrrrrr}
\tabletypesize{\scriptsize}
\tablecaption{Mean errors of parameters with various trigonometric parallax errors $\delta_{p,{\rm tri}}$. 
\label{tbl:sdevacc}}
\tablewidth{0pt}
\tablehead{
$\delta_{p,{\rm tri}}$ & $\bar{M}$ & $\sigma_{R,\sun}$ & $\phi_0$ & $c$ &  $\Omega_p$ & $\Phi_{a}$ & $h_\sigma$ & $h$\\
}
\startdata
   1 &  0.10 &  0.58 &  0.05 &  0.13 &  0.33 &  32.8 &  0.07 &  0.22\\
   2 &  0.10 &  0.57 &  0.05 &  0.14 &  0.34 &  33.5 &  0.08 &  0.26\\
   4 &  0.11 &  0.58 &  0.05 &  0.14 &  0.40 &  37.3 &  0.08 &  0.28\\
   8 &  0.12 &  0.59 &  0.06 &  0.16 &  0.41 &  37.3 &  0.09 &  0.29\\
  10 &  0.13 &  0.60 &  0.06 &  0.17 &  0.41 &  36.7 &  0.09 &  0.29\\
  20 &  0.14 &  0.61 &  0.06 &  0.18 &  0.43 &  38.2 &  0.09 &  0.32\\
  50 &  0.15 &  0.63 &  0.06 &  0.18 &  0.44 &  38.2 &  0.10 &  0.35\\
 100 &  0.15 &  0.65 &  0.06 &  0.18 &  0.45 &  39.0 &  0.10 &  0.37\\
 200 &  0.16 &  0.65 &  0.06 &  0.19 &  0.45 &  38.7 &  0.11 &  0.38\\
 500 &  0.18 &  0.67 &  0.06 &  0.20 &  0.47 &  40.3 &  0.12 &  0.40\\
 1000 &  0.23 &  0.68 &  0.07 &  0.22 &  0.50 &  41.8 &  0.14 &  0.54\\
 2000 &  0.31 &  0.74 &  0.07 &  0.23 &  0.56 &  46.1 &  0.15 &  0.57\\
\enddata
\tablecomments{
For $N=850$ and fixing $R_0=8$ kpc.  See the notes for Table \ref{tbl:sdevN}.
}
\end{deluxetable}

\begin{deluxetable}{rrrrrrrrrrr}
\tabletypesize{\scriptsize}
\tablecaption{Mean errors of parameters with various photometric parallax errors $\delta_{p,{\rm phot}}$.
\label{tbl:sdevpphot}}
\tablewidth{0pt}
\tablehead{
$\delta_{p,{\rm tri}}$ & $\delta_{p,{\rm phot}}$ & $\bar{M}$ & $\sigma_{R,\sun}$ & $\phi_0$ & $c$ &  $\Omega_p$ & $\Phi_{a}$ & $h_\sigma$ & $h$\\
}
\startdata
10 & 0.10 &  0.14 &  0.60 &  0.06 &  0.17  &  0.48 &  42.0 &  0.09 &  0.45\\
10 & 0.15 &  0.13 &  0.60 &  0.06 &  0.17  &  0.41 &  36.7 &  0.09 &  0.29\\
10 & 0.20 &  0.12 &  0.59 &  0.05 &  0.15  &  0.39 &  37.6 &  0.09 &  0.23\\
1000& 0.10 &  0.23 &  0.66 &  0.06 &  0.21  &  0.51 &  40.5 &  0.12 &  1.07\\
1000& 0.15 &  0.23 &  0.68 &  0.07 &  0.22  &  0.50 &  41.8 &  0.14 &  0.54\\
1000& 0.20 &  0.23 &  0.70 &  0.06 &  0.20  &  0.47 &  41.0 &  0.14 &  0.37\\
\enddata
\tablecomments{
For $N=850$ and fixing $R_0=8$ kpc. $\delta_{p,{\rm tri}}$ is in $\mu$as. See the notes for Table \ref{tbl:sdevN}.
}
\end{deluxetable}

\begin{deluxetable}{rrrrrrrrrrr}
\tabletypesize{\scriptsize}
\tablecaption{Mean errors of parameters with various angular coverages $\phi_{\rm max}$. 
\label{tbl:sdevphimax}}
\tablewidth{0pt}
\tablehead{
$N$ & $\phi_{\rm max}$ & $\bar{M}$ & $\sigma_{R,\sun}$ & $c$ & $\phi_0$ &  $\Omega_p$ & $\Phi_{a}$ & $h_\sigma$ & $h$ & $R_0$ \\
}
\startdata
850 & 133 &  0.13 &  0.60 &  0.06 &  0.17 &  0.41 &  36.7 &  0.09 &  0.29 &    ---\\
850 &  90 &  0.13 &  0.61 &  0.06 &  0.17 &  0.41 &  37.6 &  0.09 &  0.29 &    ---\\
850 &  60 &  0.14 &  0.58 &  0.05 &  0.17 &  0.39 &  37.4 &  0.09 &  0.36 &    ---\\
850 &  30 &  0.18 &  0.59 &  0.06 &  0.20 &  0.37 &  35.6 &  0.08 &  0.48 &    ---\\
850 &  20 &  0.22 &  0.58 &  0.07 &  0.22 &  0.42 &  39.8 &  0.08 &  0.46 &    ---\\
850 &  10 &  0.40 &  0.58 &  0.06 &  0.20 &  0.45 &  43.4 &  0.08 &  0.57 &    ---\\
\hline
850 & 133 &  0.21 &  0.66 &  0.07 &  0.17 &  0.42 &  37.5 &  0.09 &  0.32 &   0.12\\
850 &  90 &  0.25 &  0.68 &  0.08 &  0.17 &  0.41 &  37.7 &  0.09 &  0.33 &   0.16\\
850 &  60 &  0.37 &  0.72 &  0.11 &  0.19 &  0.41 &  39.6 &  0.09 &  0.38 &   0.24\\
850 &  30 &  0.79 &  1.15 &  0.24 &  0.34 &  0.44 &  41.4 &  0.08 &  0.40 &   0.56\\
\hline
2000 & 133 &  0.08 &  0.39 &  0.04 &  0.10 &  0.30 &  27.5 &  0.06 &  0.19 &    ---\\
2000 &  90 &  0.08 &  0.39 &  0.04 &  0.11 &  0.28 &  26.7 &  0.06 &  0.21 &    ---\\
2000 &  60 &  0.09 &  0.39 &  0.04 &  0.12 &  0.28 &  26.5 &  0.06 &  0.23 &    ---\\
2000 &  30 &  0.12 &  0.39 &  0.04 &  0.13 &  0.30 &  29.5 &  0.06 &  0.25 &    ---\\
2000 &  20 &  0.15 &  0.37 &  0.04 &  0.13 &  0.29 &  29.1 &  0.05 &  0.27 &    ---\\
2000 &  10 &  0.25 &  0.38 &  0.04 &  0.14 &  0.31 &  29.6 &  0.05 &  0.29 &    ---\\
\hline
2000 & 133 &  0.13 &  0.42 &  0.05 &  0.11 &  0.30 &  27.2 &  0.06 &  0.19 &   0.08\\
2000 &  90 &  0.16 &  0.42 &  0.05 &  0.11 &  0.26 &  24.5 &  0.06 &  0.23 &   0.10\\
2000 &  60 &  0.23 &  0.59 &  0.07 &  0.13 &  0.28 &  28.0 &  0.06 &  0.26 &   0.16\\
2000 &  30 &  0.66 &  0.61 &  0.22 &  0.31 &  0.34 &  30.6 &  0.05 &  0.27 &   0.49\\
2000 &  20 &  0.92 &  0.61 &  0.29 &  0.38 &  0.36 &  33.6 &  0.05 &  0.31 &   0.67\\
\enddata
\tablecomments{
The case of $\delta_{p,{\rm tri}}=10$ $\mu$as. MCMC fails to converge for 
$\phi_{\rm max}\le20^\circ$ and $\le 10^\circ$ for $N=850$ and $2000$, respectively,  
in the case where $R_0$ is fitted.  ``$-$" means that $R_0$ is fixed.
See the notes for Table \ref{tbl:sdevN}.
}
\end{deluxetable}

\begin{deluxetable}{rrrrrrrrr}
\tabletypesize{\scriptsize}
\tablecaption{Number of observable stars by SIM Lite in 240 hr with various extinctions and limiting magnitude.
\label{tbl:NobsSIM}}
\tablewidth{0pt}
\tablehead{
$\delta_{p,{\rm tri}}$ & $ \underline{A_{V, \rm win}}$ & \multicolumn{7}{c}{$N(V_{\rm l} [{\rm mag]})$}  \\
$(\mu {\rm as})$ & $A_{V,\rm typ}$  & $V_{\rm l}$=$20$ & $V_{\rm l}$=19 & $V_{\rm l}$=18 & $V_{\rm l}$=17
& $V_{\rm l}$=16 &  $V_{\rm l}$=15 & $V_{\rm l}$=14\\
}
\startdata
8   & 1.0 &  24  &  --- & ---  & --- & --- & ---  & --- \\
8   & 0.8 &  37  &   70 & ---  & --- & --- & ---  & --- \\
8   & 0.5 &  72  &  129 & 213  & 336 & --- & ---  & --- \\
8   & 0.3 & 127  &  213 & 323  & 458 & 631 &  833 & --- \\
8   & 0.0 & 177  &  360 & 566  & 777 & 957 & 1069 & 1131\\
\hline
10  & 1.0 &  45  &  --- & ---  & --- & ---  & ---  & --- \\
10  & 0.8 &  69  &  128 & ---  & --- & ---  & ---  & --- \\
10  & 0.5 &  132 &  225 & 354  & 521 & ---  & ---  & --- \\
10  & 0.3 &  224 &  355 & 505  & 669 &  853 & 1031 & --- \\
10  & 0.0 &  304 &  550 & 779  & 969 & 1108 & 1182 & 1215\\
\hline
20  & 1.0 & 119  &  --- & ---  & ---  & --- & ---  & --- \\
20  & 0.8 & 172  &  296 & ---  & ---  & --- & ---  & --- \\
20  & 0.5 & 305  &  470 & 656  & 864  & --- & ---  & --- \\
20  & 0.3 & 474  &  664 & 835  & 996  & 1126& 1212 & --- \\
20  & 0.0 & 584  &  856 & 1030 & 1147 & 1211& 1235 & 1237\\
\hline
100 & 1.0 &  295 &  --- & ---  & ---  & ---  & ---  & --- \\
100 & 0.8 &  399 &  594 & ---  & ---  & ---  & ---  & --- \\
100 & 0.5 &  613 &  807 & 974  & 1108 & ---  & ---  & --- \\
100 & 0.3 &  808 &  969 & 1083 & 1163 & 1209 & 1233 & --- \\
100 & 0.0 &  900 & 1079 & 1166 & 1212 & 1231 & 1237 & 1237\\
\enddata
\tablecomments{
The numbers of observable stars $N(V_{\rm l} [{\rm mag]})$ are estimated by using the required 
mission time for SIM Lite given by Figure \ref{fig:missiontime}.
A factor $A_{V,\rm win}/{A_{V,\rm typ}}$ indicates the fraction of $V$-band extinction at low extinction
window relative to the typical extinction in that direction (see Section \ref{sec:sim}).
$V_{\rm l}$ represents $V$-band limiting magnitude to select sample stars.
``---" means that stars can not be sampled in all $R_i$ bins with that $V_{\rm l}$.
}
\end{deluxetable}

\begin{deluxetable}{rrrrrrrrr}
\tabletypesize{\scriptsize}
\tablecaption{$\bar{\delta}_M$ by SIM Lite in 240 hr with various extinctions and limiting magnitude.
\label{tbl:deltaMSIM}}
\tablewidth{0pt}
\tablehead{
$\delta_{p,{\rm tri}}$ & $ \underline{A_{V, \rm win}}$ & \multicolumn{4}{c}{$\bar{\delta}_M$}  \\
$(\mu {\rm as})$ & $A_{V,\rm typ}$  & $V_{\rm l}$=$20$ & $V_{\rm l}$=19 & $V_{\rm l}$=18 & $V_{\rm l}$=17 \\
}
\startdata
100 & 1.0 &  0.311 &  ---   & ---    & ---  \\
100 & 0.8 &  0.238 &  0.216 & ---    & ---  \\
100 & 0.5 &  0.178 &  0.161 & 0.149  & 0.156\\
\enddata
\tablecomments{
$\bar{\delta}_M$ are estimated for the cases given in Table \ref{tbl:NobsSIM} 
with $\delta_{p,{\rm tri}}=100$\,$\mu$as and fixing $R_0$.
The survey regions are as shown in Figure \ref{fig:plotxy-AV}.
The numbers of sample stars are as in Table \ref{tbl:NobsSIM}.
}
\end{deluxetable}

\clearpage

\end{document}